\documentclass[12pt]{article}
\usepackage{epsfig,amsfonts,amssymb}
\usepackage{hyperref}
\usepackage[numbers, square, comma, sort&compress]{natbib}
\def\ZZZ{{\hbox{ Z\kern-1.6mm Z}}}
\def\RRR{{\hbox{ R\kern-2.4mm R}}}
\def\CCC{{\hbox{ C\kern-2.0mm C}}}
\def\zzz{{\hbox{z\kern-1mm z}}}

\newcommand{\nn}{\nonumber \\}

\newcommand{\qeq}{{\hbox{=\kern-2.3mm ? \kern.5mm }}}
\renewcommand{\qeq}{=}

\newcommand{\eps}{\epsilon}

\newcommand{\wJ}{\wt J}

\newcommand{\wt}{\widetilde}
\newcommand{\wh}{\widehat}

\newcommand{\be}{\begin{equation}}
\newcommand{\ee}{\end{equation}}
\newcommand{\ben}{\begin{eqnarray}\displaystyle}
\newcommand{\een}{\end{eqnarray}}

\newcommand{\refb}[1]{(\ref{#1})}
\newcommand{\p}{\partial}
\newcommand{\sectiono}[1]{\section{#1}\setcounter{equation}{0}}

\def\one{{\hbox{ 1\kern-.8mm l}}}
\def\zero{{\hbox{ 0\kern-1.5mm 0}}}

\begin{document}

\baselineskip 24pt
\begin{flushright}
  HRI/ST/0916
\end{flushright}
\vspace{5mm}

\begin{center}
{\Large \bf Black Hole Hair Removal: Non-linear Analysis
}

\end{center}

\vskip .6cm
\medskip

\vspace*{4.0ex}

\baselineskip=18pt

\centerline{\large \rm   Dileep P. Jatkar, Ashoke Sen and
Yogesh K. Srivastava}

\vspace*{4.0ex}

\centerline{\large \it Harish-Chandra Research Institute}

\centerline{\large \it  Chhatnag Road, Jhusi,
Allahabad 211019, INDIA}

\vspace*{1.0ex}
\centerline{E-mail: dileep, sen, yogesh@mri.ernet.in }

\vspace*{5.0ex}

\centerline{\bf Abstract} \bigskip

BMPV black holes in flat transverse space and in Taub-NUT space
have identical near horizon geometries but different
microscopic degeneracies. It has been proposed that this difference
can be accounted for by different contribution to the degeneracies
of these black holes from hair modes, -- degrees of freedom living
outside the horizon. In this paper we explicitly construct the
hair modes of these two black holes as finite bosonic and
fermionic deformations of the black hole solution satisfying the full
non-linear equations of motion of supergravity and
 preserving the supersymmetry of the original solutions.
Special care is taken to ensure that
these solutions do not have any curvature singularity at the future
horizon when viewed as the full ten dimensional geometry.
We show that after removing the contribution due to the hair degrees
of freedom from the microscopic partition function, the partition
functions of the two black holes agree.

\vfill \eject

\baselineskip=18pt

\tableofcontents

\sectiono{Introduction} \label{s1}

String theory has been successful in providing an explanation of the
entropy of supersymmetric extremal black holes in terms of microscopic
degrees of freedom.  Initial studies focussed on black holes carrying
large charges for which the classical two 
derivative action, and the
associated formula for the entropy due to Bekenstein and Hawking, is
sufficient to compute the entropy.  This assumption can be relaxed to
some extent using Wald's formula for black hole
entropy\cite{9307038,9312023,9403028,9502009} that takes into account
higher derivative corrections to the classical action. However a
complete expression for the entropy of a black hole receives
contribution 
from higher derivative corrections as well as quantum 
corrections. On general grounds
one would expect that the generalization of Wald's formula to the full
quantum theory will involve some computation in string theory on the
near horizon geometry of the black hole and will not be sensitive to
the nature of the solution away from the horizon\cite{0809.3304}.
Indeed Wald's classical formula for the entropy certainly satisfies
this criterion.

This simple assumption has a non-trivial consequence: two
different black holes with identical near horizon geometries 
have the same macroscopic entropy. The equality of the
macroscopic and the microscopic entropy would then imply that
they must have the same microscopic entropy. There is however
a counterexample: a rotating black hole in type IIB string theory
compactified on $K3\times S^1$, known as the BMPV
black hole\cite{9602065}, placed in a flat transverse
space and in Taub-NUT space\cite{0209114}
have identical near horizon geometries\cite{0503217}
but different microscopic
degeneracies\cite{9608096,9607026,0412287,0505094,0506249,
0605210,0708.1270}!

The following resolution to this puzzle was proposed in
\cite{0901.0359}. Whereas an appropriate computation in string theory
in the near horizon geometry of the black hole would give the
macroscopic entropy associated with the horizon, the full
macroscopic entropy also involves
contribution from the hair degrees of freedom -- degrees
of freedom living outside the horizon. For a supersymmetric
black hole the latter
can be computed by identifying classical
supersymmetry preserving
normalizable deformations\footnote{The
deformations we shall   consider will always be along a null vector
or tensor and hence the norm will vanish identically. We shall
call a deformation normalizable if it vanishes at infinity and produces
a configuration with finite ADM mass / charge.} 
of the black hole solution with
support outside the horizon, and then carrying out geometric
quantization on the space of these solutions. Ref.\cite{0901.0359}
identified a class of such deformations both for the BMPV
black hole in flat transverse space and BMPV black hole in
Taub-NUT space and found that after removing the contribution from
these hair degrees of freedom from the microscopic degeneracy
formul\ae, one obtains identical result for the two black holes.
This can then be identified as the common
contribution to the degeneracy
coming from
the horizon.

The purpose of this paper is to fill some of the gaps in the analysis
of \cite{0901.0359}. These are of three types:
\begin{enumerate}
\item Ref.\cite{0901.0359} identified the bosonic deformations of the black
hole solution by working with the linearized equations of motion.
We extend them to the solutions to full non-linear equations of
motion.

\item Ref.\cite{0901.0359} gave a general argument for the existence of
a certain set of fermionic deformations but did not construct
them explicitly. We construct these fermionic modes by
solving the equations of motion of the fermions around the
BMPV black hole background.

\item Ref.\cite{0901.0359} did not study supersymmetry properties of the
deformations explicitly. We demonstrate that the deformations
preserve the same number of supersymmetries as the original
BMPV black hole background.
\end{enumerate}
During this investigation we also found an unexpected result: one set
of deformations for each black hole have mild curvature singularities
in the future   horizon when viewed as ten dimensional
geometries\cite{9612248,9701077}.
This forces us to remove these modes from the counting of the hair degrees
of freedom. Fortunately however they give identical contribution to the
partition function for both black holes and hence even after removing
their contribution from the hair partition function, we continue to
get agreement between the partition functions of the two black holes
after hair removal.

In order to guide the reader through the rest of the paper we shall
now briefly list the hair modes of both types of black holes
which we shall construct.
Since the solution is independent of the coordinate along $S^1$ it
is often useful to regard this as a string like object extended along
$S^1$. In this case
a left-moving mode will represent
a set of deformations labelled by an arbitrary
function of the light-cone
coordinate that describes propagation of a plane wave along
the negative $S^1$ direction.
We begin with BMPV black hole in
flat transverse space. In this case the hair modes are expected
to consist of
{\it (i)} four  left-moving bosonic modes describing
the transverse oscillations of the black string and
{\it (ii)} four  left-moving fermionic modes describing
propagation of the goldstino modes associated with some
broken supersymmetries.
On the other hand BMPV black hole in transverse Taub-NUT space
is expected to carry {\it (i)} three   left-moving bosonic modes
describing the oscillation of the black string in three transverse
directions,\footnote{Asymptotically Taub-NUT has the form of
$\wt S^1\times \RRR^3$; thus there are three transverse directions.}
{\it (ii)} 21   left-moving bosonic modes arising from certain
oscillation modes of the 2-form fields, {\it (iii)} four left-moving
fermionic modes describing
propagation of the goldstino modes associated with some
broken supersymmetries, and {\it (iv)} four more left-moving
bosonic modes describing the transverse oscillation of the
BMPV black string {\it relative to the} Taub-NUT space.
We  explicitly construct each of these modes in our
analysis.\footnote{In
this context we would like to mention that since most of our argument
towards the absence (or triviality) of higher order corrections to
the solution
is due to our inability to contract indices, and do not need to make
explicit use of the form of the action, 
we expect these deformations 
to
survive even after inclusion of higher derivative corrections.}
\S\ref{s2} is devoted to the construction of the hair modes
of BMPV black hole in flat transverse space, \S\ref{s3}
contains the construction of the hair modes of BMPV black hole
in Taub-NUT space and \S\ref{s4}  contains a proof that the
modes constructed in \S\ref{s2} and \S\ref{s3} preserve all
the supersymmetries of the undeformed background.
However we show
in appendix \ref{sappc} following
\cite{9612248,9701077,9605224,9606113} that the four
bosonic modes describing
the transverse oscillations of the black string in flat transverse
space and the four   bosonic modes describing
the transverse oscillations of the black string relative to the Taub-NUT
space have mild curvature singularity at the future even horizon,
Thus they should not be counted as hair degrees of freedom.
In \S\ref{s5} we compute the partition function associated with
the horizons of the two black holes by dividing the microscopic
partition function by the partition function associated with the
hair and show that the results match.

Finally we note that besides the hair modes described above, both black
holes carry twelve fermionic zero modes associated with the broken
supersymmetry generators. The construction of these zero modes is
straightforward\cite{9505116}; we take a local
supersymmetry transformation whose parameter approaches a constant
spinor other than the Killing spinor at infinity and vanishes at the
horizon, and apply it to the original black hole
solution to generate a fermionic zero mode.
Since there are 12 independent supersymmetry
transformations whose parameters do not approach a Killing spinor
at infinity, this generates 12 fermion zero modes. We shall not
discuss the construction of these zero modes any further, but count
them in computing the partition function of the hair modes in
\S\ref{s5}.

\sectiono{BMPV Black Hole Hair} \label{s2}

In this section we shall analyze the deformations of the BMPV black
hole representing its hair modes, \i.e.\ deformations which live
outside the horizon and do not change the near horizon geometry. The
theory that we shall study is type IIB supergravity compactified on
$K3$\cite{romans,9712176,9804166}. The effective six dimensional
theory of massless fields that one gets has many fields but we shall
list only those which will play a role in our analysis. We denote by
$\Phi$ the ten dimensional dilaton, by $G_{MN}$ ($0\le M,N\le 5$)
the string metric in six dimensions, by $C^{(2)}$ the RR 2-form
field and by $F^{(3)}=d\, C^{(2)}$ the associated field strength.
The theory also has several other 2-form fields. One of them comes
from  the NSNS sector and has no constraint on its field strength,
but there are 22 others obtained by dimensional reduction of the RR
4-form on 2-cycles of K3, of which 19 have anti-self-dual field
strength and 3 have self-dual field strength. Including the RR
2-form field $C^{(2)}$ of the ten dimensional theory, we have
altogether 21  2-form fields with anti-self-dual field strength and
5 2-form fields with self-dual field strength. We shall denote the
self-dual and the anti-self-dual field strengths by $\bar H^k_{MNP}$
($1\le k\le 5$) and $H^s_{MNP}$ ($6\le s\le 26)$ respectively,
satisfying
\be \label{eself} \bar H^{kMNP} = {1\over 3!} \, |\det
g|^{-1/2} \eps^{MNPQRS}\, \bar H^k_{QRS}, \qquad H^{sMNP} = -{1\over
3!} \, |\det g|^{-1/2} \eps^{MNPQRS}\, H^s_{QRS} \, ,
\ee
where $\eps^{MNPQRS}$ is the totally anti-symmetric symbol. We shall
describe our choice of the sign convention for $\eps$ shortly. The
theory also contains a set of scalar fields besides the dilaton,
coming from the moduli of $K3$, the RR scalar, as well as the
components of the NSNS 2-form field and the RR 2- and 4-form
fields
along the two and four cycles of $K3$. Throughout this paper we
shall set all the scalar fields including the dilaton to fixed
values.\footnote{When the background scalar fields are not
constants, the self-dual and anti-self-dual field strengths are not
closed but can be expressed as linear combinations of closed 3-forms
with coefficients given by functions of the scalar
fields\cite{romans, 9712176,9804166}. This complication is absent
when the scalars are constants in space-time.} The fermion fields in
this six dimensional theory consist of a set of four left-chiral
gravitini $\Psi_\mu^\alpha$ ($0\le \mu\le 5$, $1\le\alpha\le 4$) and
a set of $4\times 21$ right-chiral spin 1/2 fermions $\chi^{\alpha
r}$ ($1\le r\le 21$). 
The precise form of the chirality projection rules is given in
\refb{egravchiral}, 
\refb{echichiral}.
Note that we have suppressed the Dirac
indices.

The field strengths $\bar H^k_{MNP}$ and $H^s_{MNP}$
will  include the self-dual and
anti-self-dual parts of $F^{(3)}$.
We shall choose the convention where $\bar H^1$ and $H^6$ denote
the self-dual and anti-self-dual components of $F^{(3)}$ up to
a normalization. More precisely we choose
\be \label{ef3}
F^{(3)}_{MNP} = 2\, e^{-\Phi}\,
\left(\bar H^1_{MNP} + H^6_{MNP}\right)\, ,
\ee
where $\Phi$ is the constant value of the dilaton field.
The
self-dual-field strengths $\bar H^2,\cdots \bar H^5$ will be set to zero
and will play no role  throughout our
analysis.
In the sector where
all the scalar fields are constants and fermions are set to zero,
the bosonic
equations of motion
take the form:\footnote{This requires appropriate normalization
factors appearing in the definition of the $\bar H^k$'s and $H^s$'s in terms
of the fundamental fields of string theory. Typically these normalization
factors will be functions of the various scalar fields in six dimensions
but as long as the scalar fields are frozen to constant values we do not
have to worry about these normalizations.}
\ben \label{eeom}
& R_{MN} = \bar H^k_{MPQ} \, \bar H_N^{kPQ} +
H^s_{MPQ} \, H_N^{sPQ}\, \nn
& \bar H^k_{MNP} H^{sMNP} = 0\, ,  
\een
where $R_{MN}$ is the Ricci tensor defined in the
sign convention
in which on the sphere the Ricci scalar $G^{MN}R_{MN}$
is positive.

We now further compactify the theory on $S^1$ and
consider a rotating black hole solution describing
$Q_5$ D5-branes along
$K3\times S^1$, $Q_1$ D1-branes along $S^1$,
$-n$ units of momentum along $S^1$ and angular momentum
$J$\cite{9602065}.
We denote by
$x^5$ the coordinate of the circle
$S^1$ with period $2\,\pi\, R_5$,
by ${\left( 2\pi \right)^4 V}$
the volume of $K3$
measured in the string metric, and by
$\lambda$ the asymptotic value of the string coupling.
As in \cite{0901.0359} we shall set the  asymptotic values of the scalar
fields to their attractor values to keep the solution simple.
We also denote by $t$ the time coordinate and by $w_i$ ($1\le i\le 4$)
the four non-compact spatial coordinates. Finally
we denote by $(r,\theta,\phi, x^4)$ the Gibbons-Hawking coordinates
of the four dimensional space labelled by $\vec w$ so that we have
\ben \label{ecart1}
w^1 = 2\sqrt r \cos{\theta\over 2} \cos{x^4+\phi\over 2},
&& w^2 = 2\sqrt r \cos{\theta\over 2} \sin{x^4+\phi\over 2},
\nonumber \\
w^3 = 2\sqrt r \sin{\theta\over 2} \cos{x^4-\phi\over 2},
&& w^4 = 2\sqrt r \sin{\theta\over 2} \sin{x^4-\phi\over 2}\, ,
\nonumber \\
(\theta,\phi, x^4)\equiv(2\pi-\theta, \phi+\pi, x^4+\pi)
&\equiv& (\theta, \phi+2\pi,x^4+2\pi) \equiv (\theta,\phi,x^4+4\pi)\, .
\een
In this case we have
\ben \label{egibb}
&& r = {1\over 4}\,
w^i w^i\, , \quad
 dw^i dw^i  =    r\,
(dx^4 +\cos\theta d\phi)^2 +  {1\over r}  \,
(dr^2 + r^2 d\theta^2 + r^2\, \sin^2\theta \, d\phi^2)\, ,
\een
and
the solution takes the form
\ben \label{ep4}
ds^2 &\equiv &  G_{MN} dx^M dx^N \nonumber \\
&=& \psi^{-1}(r)
 \left[ du\, dv + \left(\psi(r)-1\right) dv^2
 + \chi_i (r) \, dv \, dw^i
 \right]
+\psi(r)dw_i dw_i
\nonumber \\
e^{-2\Phi}&=&\lambda^{-2} \, , \nonumber \\
F^{(3)}&\equiv&
{1\over 6} \, F^{(3)}_{MNP} dx^M \wedge dx^N\wedge
dx^P = {r_0\over \lambda}\,
\left( \epsilon_3 +  *_6 \epsilon_3
+{1\over r_0} \psi^{-1}(r)
dv\wedge d \zeta\right) \, ,
\een
where $G_{MN}$ is the six dimensional string metric and
\ben \label{edefs}
u&\equiv& x^5-t, \quad v \equiv x^5+t,  \nonumber \\
\psi(r) &\equiv& \left( 1 + { r_0  \over r } \right)\, ,
\nonumber \\
 \chi_i \, dw^i
&\equiv& -2\, \zeta\, , \quad
 \zeta\equiv  -{\wJ \over 8r} \, (dx^4 +\cos\theta d\phi)\, , \nonumber \\
\eps_3 &\equiv&  \sin\theta\, dx^4 \wedge d\theta\wedge d\phi\, . 
\een
Here
$*_6$ denotes Hodge
dual in the six dimensions spanned by $t$, $x^5$, $x^4$,
$r$, $\theta$
and $\phi$ with the convention $\epsilon^{t54r\theta\phi}=1$.
The constants $r_0$ and $\wt J$
are given in terms of the
charges and the asymptotic values of the moduli fields as follows:
\begin{equation} \label{ep5}
r_0  = {\lambda (Q_1-Q_5)
 \over 4V }  = {\lambda Q_5  \over 4
} = { \lambda^2 |n |  \over 4R_5^2 V }\, , \qquad
\wJ = {J\, \lambda^2\over 2\, R_5\, V}\, .
\end{equation}
Eq.\refb{ep5} gives specific relations between $V$, $\lambda$
and $R_5$
 reflecting the fact we have chosen them to
coincide with the attractor values instead of keeping them general.
For later use we note that the background metric and the three
form field strengths can be expressed as
\ben \label{enewform}
ds^2 &=& - (e^0)^2 + (e^1)^2 + (e^2)^2 + (e^3)^2
+(e^4)^2 + (e^5)^2
\, ,\nonumber \\
F^{(3)} &=& {r_0 \over \lambda\, r^2}\,
\left[ \psi^{-3/2}(r)\,
r^{1/2}\, (e^2\wedge e^4\wedge e^5
+ e^0\wedge e^1\wedge e^3) \right. \nonumber \\ &&
\left.
+ {\wJ \over 8\, r_0} \, \psi^{-2}(r)\,
(-e^0\wedge e^2\wedge e^3 + e^0\wedge e^4\wedge e^5
-e^1\wedge e^2\wedge e^3 + e^1\wedge e^4\wedge e^5)
\right]\, , \nonumber \\
\een
where
\ben \label{eoneform}
e^0 &=& \psi^{-1}(r)\, (dt+ \zeta), \nonumber \\
e^1 &=& \left(dx^5+dt- \psi^{-1}(r)
(dt+ \zeta) \right),   \nonumber \\
e^2 &=&
\psi^{1/2}(r) \, r^{1/2} (dx^4 +\cos\theta d\phi),
\nonumber \\
e^3 &=&
\psi^{1/2}(r) \, r^{-1/2} \, dr
\, , \nonumber \\
e^4 &=&  \psi^{1/2}(r)\, r^{1/2}  \, d\theta
\, , \nonumber \\
e^5 &=& \psi^{1/2}(r)\, r^{1/2}  \,
\sin\theta\, d\phi\, .
\een
The one
forms $e^A$ are related to the vielbeins $e^{~A}_M$ via the
relations
\be \label{eviel}
e^A = e^{~A}_M dx^M\, .
\ee
Here $A$ labels a tangent space index.
{}From \refb{enewform} it follows that
the fields strength $F^{(3)}$ appearing in \refb{ep4} is self-dual.
Thus in the black hole background all the
anti-self-dual field strengths $H^s_{MNP}$'s vanish.

The near horizon geometry of \refb{ep4} is obtained by introducing
new coordinates $\rho$, $\tau$ via:
\be \label{ep8} r= r_0\,
\beta\rho, \quad t= \tau/\beta,
\ee
and taking the limit $\beta\to
0$ keeping $\tau$, $v$, $\rho$, $x^4$, $\theta$ and $\phi$ finite.
In this limit \refb{ep4}  takes the form
\ben \label{ep9}
ds^2&=&r_0
{d\rho^2\over \rho^2} + dv^2 +r_0(dx^4 + \cos\theta d\phi)^2 +{\wJ
\over 4 r_0} dv (dx^4 + \cos\theta d\phi) - 2\rho\, dv d\tau
 \nonumber \\ &&  +r_0  \left(
 d\theta^2 + \sin^2\theta d\phi^2\right)
\, , \nonumber \\
e^{\Phi} &=& \lambda\, , \nonumber \\
F^{(3)} &=& {r_0\over \lambda}\, \left[\eps_3 + *\eps_3
+{\wJ \over 8\, r_0^2}\, dv\wedge \left({1\over \rho}\, d\rho\wedge
(d \, x^4 +\cos\theta\, d\phi)
+ \sin\theta\, d\theta\wedge d\phi\right)\right]\, .
\een

We shall now analyze various bosonic and fermionic deformations
of this solution which live outside the horizon.
This in particular will require that 
when expressed in terms of the
new coordinate system \refb{ep8} the deformations should vanish as
$\beta\to 0$.
Geometric quantization of these deformations are
supposed to generate the degeneracies associated with the hair
modes. The bosonic deformations representing transverse oscillation
of the black hole were constructed in \cite{0901.0359} at the linearized
level. Here however we shall go beyond the linearized approximation
and construct the fully backreacted solution.

\subsection{Bosonic deformations  representing transverse oscillation
of the BMPV black hole} \label{s2.1}

In this section we shall follow
\cite{vachaspati,9511064,9604134,9609084,9512031} to construct
deformations describing left-moving transverse oscillations
of the black hole. Even though these deformations will turn out
to be singular at the future horizon\cite{9612248,9701077} and hence
will not be counted among the hair degrees of freedom, we shall
go through it carefully as similar deformations of the four dimensional
solution will turn out to be non-singular and hence will correspond
to hair degrees of freedom.

Given a space-time with metric $G_{MN}$ satisfying the supergravity
equations and a null, killing and hypersurface orthogonal vector field
$k_M$, {\it i.e.}, satisfying the following properties
\begin{equation}
  \label{eq:2}
 \  k^Mk_M = 0,\  k_{M ; N}+ k_{N;M} =0,\ k_{M;N}=
 \frac{1}{2} (k_{M}A_{,N}-k_{N}A_{,M})
\end{equation}
for some scalar function $A$,  one can construct a new exact solution
of the equations of motion by
defining\cite{vachaspati}
\begin{equation}
  \label{eq:3}
G'_{MN}= G_{MN} + e^{-A}T k_{M}k_{N}
\end{equation}
where the function $T$ satisfies
\begin{equation}
  \label{eq:4}
\nabla^{2}T =0 \ , \ k^{M}\partial_{M}T =0\, .
\end{equation}
The new metric $G'_{MN}$ describes a gravitational wave on the
background of the original metric provided the matter fields, if
any, satisfy some conditions.
We take $(\frac{\partial}{\partial u})$ 
as our null killing 
vector.  Since $G_{uu}=0$, it is obviously null and since the metric
coefficients do not depend on $u$, it is also killing.
For our case, only non-zero
component of killing one-form is $k_{v}= G_{uv}=\psi^{-1}$ and
 the hypersurface-orthogonality condition
(the last equation in \refb{eq:2})  is satisfied by choosing
$e^{-A}= \psi$.
Applying
the transform we get\cite{9511064,9604134,9609084}
\be
  \label{eq:5}
  ds^2= \psi^{-1}(r)
  \left\{dudv+(\psi-1 + T(v,\vec{w}))dv^2  
  +
  \chi_i(r)dvdw_i\right\}  
  +\psi(r)dw_i dw_i 
\ee
where $T(v,\vec{w})$ satisfies the flat four dimensional Laplace
equation:
\be \label{elaplace}
\p_{w^i} \p_{w^i} T(v, \vec w) = 0\, .
\ee
It also follows from the analysis of \cite{vachaspati} that we do not
need to modify the dilaton 
and the 2-form fields. A simple way to see
this is as follows.
For any component of a
covariant tensor, we define the weight of that component as the
number of $v$ indices minus the number of $u$ indices carried
by the tensor. For a component of the
contravariant tensor we define the weight
to be the number of $u$ indices minus the number of $v$ indices.
Then any tensor can be decomposed as a sum of tensors of
fixed weights and in the contraction of covariant and contravariant
indices the weight is preserved. Now by examining the
background \refb{ep4} we see that each term in the solution has
weight zero or positive.
On the other hand the term proportional to $T(v,\vec w)$ in
\refb{eq:5} has weight 2.
Furthermore the original background as well as the deformation
generated by $T(v,\vec w)$ are $u$ independent; hence we
cannot reduce the weight by taking $u$ derivative of the
background.
Thus the term proportional to $T(v,\vec w)$
can only produce terms in the equation of
motion of weight two or more.
In
other words it can only generate terms for which
the number of covariant $v$ indices is larger then the number
of covariant $u$ indices by at least 2. This is impossible for
the dilaton equation of motion which carries no index. The equation
of motion for the 2-form field has two indices, but it is anti-symmetric
in these two indices. Thus it is impossible to have more than one
covariant $v$ index. The only equation of motion that can be
affected by the $T(v,\vec{w})\, dv^2$ term is the $vv$ component
of the metric equation, leading to \refb{elaplace}.

We can write down the general solution to
\refb{elaplace} as an expansion in spherical
harmonics on $S^3$, but after requiring regularity
at the origin
and at infinity and dropping terms which can be removed by coordinate
transformation, we can choose
\be \label{etchoice}
T(v,\vec{w})= \vec{f}(v)\cdot \vec{w}\, , \qquad \int_0^{2\pi R_5}
f_i(v) dv = 0\, ,  
\ee
for some arbitrary set of four functions $(f_1(v),\cdots f_4(v))$
subject to the restriction given above.  
The corresponding
metric
\be
  \label{eq:5rep}
  ds^2 = \psi^{-1}(r)\left[dudv+\left\{\psi-1 +
  \vec f(v)\cdot \vec w\right\}dv^2 +
  \chi_i(r)dvdw_i\right]
  +\psi(r)dw_i dw_i
\ee
is apparently not
asymptotically flat but can be made so by the following coordinate
transformations\footnote{Note that in order   
that the deformations preserve the asymptotic geometry, the
shifted coordinates $(u',v')$ should be identified with $(x^5\mp t)$.
Thus for example the identification under $x^5\to x^5+2\pi R_5$
will act as $(u',v')\to (u'+2\pi R_5, v'+2\pi R_5)$.}
\begin{eqnarray}
  \label{eq:7}
v&=& v' \nonumber \\
\vec{w}&=&\vec{w'} + \vec{F} \nonumber \\
u &=& u' -2\dot{F_{i}}w'_{i} - 2\dot{F_{i}}F_{i}
+ \int^{v'} \dot{F}^{2}(v'')dv''.  
\end{eqnarray}
Here $\vec{f}(v)= 2\ddot{\vec{F}}$ and
dot refers to derivative with
respect to $v$. Making this change of coordinates,
the terms in metric
change as follows
\begin{eqnarray}
  \label{eq:8}
dudv &= &du'dv' -2\dot{F_{i}}dw'_{i}dv' - \dot{F_i}\dot{F_i} \,
dv'^{2}
-2\ddot{ {F}}_i\, ( {w'}_i + {F}_i) \, dv'^{2}\nonumber \\
dv\, dw_i &=& dv'\, (dw_i' +\dot F_i\, dv') \nn
dw_{j}dw_{j} &=& dw'_{j}dw'_{j} + \dot{F_i}\dot{F_i} \,
dv'^{2}
+2\dot{F_{i}}\, dw'_{i} \, dv' \, .
\end{eqnarray}
Removing the primes, we write the above metric as
\begin{equation}
  \label{eq:9}
ds^{2} = H^{-1}dudv + H dw_{j}^{2} + A_{j}dw_{j}dv + Kdv^{2}
\end{equation}
where 
\ben
  \label{eq:10}
&& H  = 1+ \frac{4\, r_{0}}{|\vec{w}+\vec{F}(v)|^{2}} , \quad
K= 1- H^{-1} + (H-H^{-1})  \dot{F}^{2}(v)
+ H^{-1}\, \chi_j\dot{F_j}(v),  \nonumber \\
&& A_{j}= H^{-1}\chi_{j} 
+ 2\dot{F}_{j}(H-H^{-1})\, .
\een
Since $A_j\to 0$, $K\to 0$ and $H\to 1$ as $|\vec w|\to\infty$,
the metric is asymptotically flat.
Note however that this change of coordinates changes the location
of the horizon, and hence it is not apparent that the deformation lives
outside the horizon. To overcome this we shall
make the coordinate transformation that takes the form
given in \refb{eq:7} for large $r$ but which becomes
identity near the horizon. In this case the coordinates near
the horizon are the original coordinates $(v,u,\vec w)$ and
the metric takes the form given in \refb{eq:5rep}.
Since in the new coordinate system \refb{ep8}
 $\psi^{-1}(r) \vec f(v)\cdot \vec w\sim \beta^{3/2}$, the deformation
vanishes as $\beta\to 0$. Thus the deformations
generated by $T$ does not affect the near horizon geometry
of the black hole, and represent
candidates for hair degrees of freedom.

To linear order in $\vec f(v)$ the solution given in \refb{eq:5rep}
can be shown to be related by a coordinate transformation
to the deformation described in
\cite{0901.0359}
representing transverse motion of the BMPV black hole. Thus
the solution \refb{eq:5rep} represents, physically, finite amplitude
oscillations of the BMPV black hole in the transverse direction
after taking into account the backreaction of the gravitational and
other fields.

Since the deformation parameters $\vec f(v)$ transform as a
vector under the SO(4) rotation in the transverse space, we expect the
deformations to carry angular momentum. This is visible
explicitly in the additional term proportional to $\dot F_j$
appearing in the expression for $A_j$. Since this modifies
the coefficient of the $dw^j dv$ term in the asymptotic metric
given in \refb{eq:9},
the deformed configuration carries additional angular momentum besides
the one associated with the undeformed solution.

{\it
We shall however see in appendix \ref{sappc} that even though these modes
apparently vanish at the horizon, they in fact have curvature
singularities at the future horizon. Thus they should be excluded
from the counting of the hair modes.}

\subsection{Fermionic deformations  associated with the
broken supersymmetry generators of the BMPV black hole} \label{s2.2}

Since the black hole solution breaks twelve of the sixteen
space-time supersymmetries,  we expect to have twelve fermionic zero
modes living on the black hole, forming part of the black hole hair.
It was argued in \cite{0901.0359} that four of these lift to full
left-moving fields on the two dimensional world volume of the black
hole spanned by $t$ and $x^5$. In that case we should be able to
construct solutions to the equations of motion of the fermion
fields
labelled by four independent functions of $v$. We shall now
explicitly construct these solutions in the undeformed background
\refb{ep4} and then argue that the solutions remain unaffected by
the deformation described in \refb{eq:5}. We shall follow the notation
of \cite{9804166}.

The linearized equation of motion of $\Psi^\alpha_M$
and $\chi^{\alpha r}$
in the background
where all the scalars are constants and $\chi^{\alpha r}$ are set
to zero are
\ben
  \label{eq:18}
&&  \Gamma^{MNP} D_N\Psi^\alpha_P
   - \bar H^{kMNP} \Gamma_N\, \wh\Gamma^k_{\alpha\beta}
   \Psi^\beta_P= 0, \nn
&& H^{sMNP} \, \Gamma_{MN} \Psi^\alpha_P=0\, ,
\een
where
\be \label{edefcov}
D_M\Psi^\alpha_P = \p_M \Psi^\alpha_P - \Gamma^N_{MP}
\Psi^\alpha_N + {1\over 4} \,
\omega_M^{AB} \, \wt\Gamma^{AB}\,  \Psi^\alpha_P\, ,
\ee
\be \label{edefomgam}
\Gamma^M_{NP} \equiv {1\over 2}\, G^{MR}\,
(\p_N G_{PR} + \p_P G_{NR} -\p_R G_{NP}),
\qquad \omega_M^{AB} \equiv -G^{NP} e_N^{~B} \p_M
e_P^{~A} +  e_N^{~A} e_P^{~B} G^{PQ} \Gamma^N_{QM}\, .
\ee
Since in our background $F^{(3)}$ is self-dual, we have
$H^{sMNP}=0$ and hence the second set of equations in \refb{eq:18}
is automatically satisfied.
The first set of equations involves
only the self-dual part of the 3-form
denoted by $\bar H^k_{MNP}$ for $1\le k\le 5$.
In \refb{eq:18} $\Gamma^M$'s  $(0\le M\le 5)$
denote $8\times 8$ $SO(5,1)$ gamma matrices written
in the coordinate basis and  $\wh\Gamma^i$ denote the
$4\times 4$ $SO(5)$
gamma matrices, satisfying
\be \label{egpr}
\{\Gamma^M, \Gamma^N\} = 2\, G^{MN}, \qquad
\{\wh\Gamma^k, \wh\Gamma^l\} =2\, \delta_{kl}\, ,
\ee
and $\Gamma^{M_1\cdots M_k}$ is the totally
anti-symmetric product of
$\Gamma^{M_1},\cdots \Gamma^{M_k}$.
It will also be useful to introduce the gamma matrices
$\wt\Gamma^A$ carrying SO(5,1) tangent
space indices:
\be \label{etang}
\wt \Gamma^A = e^{~A}_M \Gamma^M, \qquad \{\wt\Gamma^A,
\wt\Gamma^B\}=2\, \eta^{AB}\, .
\ee
In this convention the  fields $\Psi_M^\alpha$ and $\chi^{\alpha r}$
 satisfiy  chirality projection
conditions
\be \label{egravchiral}
\left({1\over 6!} \, |\det g|^{-1/2}\,
\eps^{MNPQRS} \Gamma_{MNPQRS} +1\right)\Psi_M^\alpha
=0 \quad \to \quad (\wt \Gamma_{012345} +1) \, \Psi_M^\alpha
=0\, ,
\ee
\be \label{echichiral}
\left({1\over 6!} \, |\det g|^{-1/2}\,
\eps^{MNPQRS} \Gamma_{MNPQRS} -1\right)\chi^{\alpha r}
=0 \, .
\ee
Note that the tangent space indices are raised and lowered by the
flat metric $\eta_{AB}$.
It follows from \refb{eoneform} that
\be \label{ereln}
\wt\Gamma^0 + \wt \Gamma^1 = \Gamma^v\, .
\ee

To solve \refb{eq:18},
we make the following ansatz for the gravitino fields:
\be \label{eansatz}
\Psi^\alpha_M = 0 \quad \hbox{for} \quad M\ne v\, ,
\ee
and furthermore that $\Psi_v^\alpha$ is $u$-independent.
We  also impose
a gauge condition on $\Psi^\alpha_M$
\begin{equation} \label{egauge}
 \Gamma^M\Psi^\alpha_M =0 \quad \to \quad  \Gamma^v
 \Psi^\alpha_v=0.
\end{equation}
Using \refb{ereln} this may be expressed as
\be \label{eng}
(\wt\Gamma^0 + \wt \Gamma^1)\Psi_v^\alpha=0 \quad \to
\quad \wt\Gamma^0 \, \wt \Gamma^1\, \Psi_v^\alpha
=\Psi_v^\alpha\, .
\ee
Since the only non-vanishing component of the
gravitino is $\Psi^\alpha_v$, we see that in the convention
described below \refb{elaplace} the fermionic deformation has
weight 1. Note that we do not assign any weight to
the SO(5,1) or SO(5) spinor indices. Consider now a term in the
equation of motion that is linear in the gravitino field.
Since the
fields in the original background are all of weight $\ge 0$,
multiplying the gravitino by these fields cannot reduce the weight.
Furthermore since $\Psi_v^\alpha$ as well as all
other background fields
is $u$ independent, we cannot
reduce the weight by acting with a $u$ derivative on the gravitino.
Finally we also cannot reduce the weight by acting with a
$\Gamma^v$ on the gravitino due to
eq.\refb{egauge}.\footnote{If there are a set of other
gamma matrices between the $\Gamma^v$ and the gravitino, we
can still bring $\Gamma^v$ next to the gravitino using eq.\refb{egpr},
and none of the extra terms generated in this process  can reduce the
weight.} Thus we conclude that any term in the equation of motion that
involves at least one power of the gravitino must be of weight $\ge 1$.
This in turn shows that the only non-trivial component of the
equation of motion \refb{eq:18} is the one associated with the choice
$M=u$. For this choice \refb{eq:18} takes the form:
\be \label{eq:18.5}
\Gamma^{uiv} \left( \p_i + {1\over 4} \,
\omega_i^{AB} \, \wt\Gamma^{AB} \right) \Psi^\alpha_v
- \bar H^{kuiv} \Gamma_i \, \wh\Gamma^k_{\alpha\beta} \Psi^\beta_v
=0\, .
\ee
The above analysis also tells us that in computing the right hand side
of \refb{eq:18.5} we only need to keep terms in the background fields
of weight zero. Thus we can ignore the terms proportional to
$dv^2$ and $dv dw^i$ in the metric and the term proportional
to $\wt J$ in $F^{(3)}$. This allows us to choose the vielbeins to be
of the form:
\ben \label{esimpleviel}
&& e^0 =  {1\over 2}  (dv-\psi^{-1}du),
\qquad e^1 = {1\over 2}   (dv+
\psi^{-1}du),
\nonumber \\
&& e^2 = \psi^{1/2}r^{1/2}
(dx^4 + \cos\theta d\phi), \qquad
 e^3 = \psi^{1/2}r^{-1/2}dr,
\nonumber \\
&& e^4 = \psi^{1/2}r^{1/2}d\theta, \qquad e^5=
\psi^{1/2}r^{1/2}\sin\theta\, d\phi\, .
\een
The associated non-vanishing components of the spin connection
are given by
\ben \label{espinsimple}
&& \omega_r^{01} = -{1\over 2}\, {\psi'\over \psi}\, ,
\quad \omega_{x^4}^{23} = {1\over 2}\, {(r\psi)'\over \psi}\, ,
\quad \omega_{x^4}^{45}={1\over 2}, \quad
\omega_\phi^{23} = {1\over 2}\, {(r\psi)'\over \psi}\, \cos\theta,
\quad \omega_\phi^{24} = - {1\over 2}\, \sin\theta, \nonumber \\
&& \omega_\phi^{35} =- {1\over 2}\, {(r\psi)'\over \psi}\, \sin\theta,
\quad \omega_\phi^{45}= - {1\over 2}\, \cos\theta\, ,
\quad \omega_\theta^{25} = {1\over 2}, \quad
\omega_\theta^{34} = -{1\over 2} \, {(r\psi)'\over \psi}\, .
\een
The same argument implies that terms quadratic and higher
powers in the gravitino fields, being of weight two or more,
cannot affect the gravitino field equations. The only equation it
could possibly affect is the $vv$ component of the metric equation,
but the projection condition \refb{eng} rules this out since it makes it
impossible to construct gravitino bilinears without any spinor index
unless one uses insertion of a $\Gamma^u$ that increases the
weight further. Thus a solution to \refb{eq:18.5} will give an
exact solution to the equations of motion.

Eq.\refb{eq:18.5} can be manipulated as follows. First of all the
$\Gamma^{uiv}$ factor may be expressed as a sum of six
terms with each term containing a different arrangement of
$\Gamma^u$, $\Gamma^i$ and $\Gamma^v$. The terms where
$\Gamma^v$ is to the extreme right vanish due to \refb{egauge}.
In the other terms we can bring $\Gamma^v$ to the extreme right
using \refb{egpr} and then use \refb{egauge} again. This allows
us to reduce the $\Gamma^{uiv}$ factor to a single gamma matrix
and leads to the equation:
\be \label{egr1}
\Gamma^i \, G^{uv}\, \left( \p_i + {1\over 4} \,
\omega_i^{AB} \, \wt\Gamma^{AB} \right) \Psi^\alpha_v
-\Gamma^i \, G^{uv}\, G^{vu}\, \bar H^k_{viu}\,
\wh\Gamma^k_{\alpha\beta} \Psi^\beta_v
=0\, .
\ee
Now dropping an overall
$G^{uv}$ factor, using \refb{ef3} and
the fact that $F^{(3)}$ is self-dual, we
get
\be \label{egr3}
\Gamma^i \, \left( \p_i + {1\over 4} \,
\omega_i^{AB} \, \wt\Gamma^{AB} \right) \Psi^\alpha_v
+{\lambda\over 2} \,
G^{vu}\, F^{(3)}_{ivu}\, \Gamma^i
\wh\Gamma^1_{\alpha\beta} \Psi^\beta_v
=0\, .
\ee
These equations are written in a covariant form in the transverse
coordinates. Thus the sum over $i$ can be taken either
over  the coordinates
$(w^1,\cdots w^4)$ or over the coordinates
$(r,x^4,\theta,\phi)$.
We shall use the $(r,x^4,\theta,\phi)$ coordinates.
Using eqs.\refb{egravchiral}, \refb{eng}
we arrive at the following equation:
\ben \label{emaster}
&& \psi^{-1/2} r^{1/2} \wt\Gamma^3\,
\left(\p_r +   {\psi'\over \psi} + {1\over r}
-{1\over 2} \, {\psi'\over \psi}\, \wh \Gamma^1\right)\Psi_v
+(r\psi)^{-1/2} (\sin\theta)^{-1} \wt\Gamma^5 \p_\phi\Psi_v\nonumber
\\ &&
+ (r\psi)^{-1/2} (\wt\Gamma^2 - \cot\theta \,
\wt\Gamma^5) \,
\p_{x^4}\Psi_v
+ (r\psi)^{-1/2} \wt\Gamma^4 \left(\p_\theta +{1\over 2} \cot\theta
\right) \Psi_v =0\, .
\een

In looking for solutions to these equations we use the fact that the
gravitino
deformation we are looking for carries no $x^4$ momentum and carries
$\pm 1/2$ units of $\phi$ momentum\cite{0901.0359}.
Thus we can require
\be \label{ereq}
\p_{x^4}\Psi_v=0, \qquad \p_\phi \Psi_v = i \, m\, \Psi_v, \quad
m=\pm{1\over 2}\, .
\ee
Substituting this into \refb{emaster} we get
\ben \label{eget}
&& \psi^{-1/2} r^{1/2} \wt\Gamma^3\,
\left(\p_r +   {\psi'\over \psi} + {1\over r}
-{1\over 2} \,   {\psi'\over \psi}\, \wh \Gamma^1\right)\Psi_v
\nonumber \\
&& +{i }\, m\,
(r\psi)^{-1/2} (\sin\theta)^{-1} \wt\Gamma^5 \, \Psi_v
+ (r\psi)^{-1/2} \wt\Gamma^4 \left(\p_\theta +{1\over 2} \cot\theta
\right) \Psi_v =0\, .
\een
We shall now rewrite this equation as
\ben \label{erew1}
&& \psi^{-1/2} r^{1/2} \wt\Gamma^3\,
\left(\p_r +   {\psi'\over \psi}
-{1\over 2} \,   {\psi'\over \psi}\, \wh \Gamma^1\right)\Psi_v
\nonumber \\
&& +(r\psi)^{-1/2}  \wt\Gamma^4\left[
{i }\, m\,
(\sin\theta)^{-1} \wt\Gamma^4 \wt\Gamma^5
+\wt\Gamma^4\wt\Gamma^3
+\left(\p_\theta +{1\over 2} \cot\theta
\right) \right]\Psi_v =0\, .
\een
We shall now find solutions to this equation by separately setting to
zero the terms in the two lines. For this
we use the following representation\footnote{If we want to
append $\wt \Gamma^2$ to this list we can take
the direct product of the matrices given in \refb{erew2}
with $\sigma_3$ and represent $\wt \Gamma^2$ as
$I_2\times \sigma_1$. This construction can be easily extended
to include $\wt\Gamma^0$ and $\wt\Gamma^1$ as well but will
not affect the analysis following \refb{erew2}.}
of $\wt\Gamma^3$, $\wt\Gamma^4$ and $\wt\Gamma^5$:
\be \label{erew2}
\wt\Gamma^4 = \sigma^1, \quad \wt\Gamma^5=\sigma^2,
\quad \wt\Gamma^3=\sigma^3\, .
\ee
Setting the second line of \refb{erew1} to zero then gives
\be \label{erew3}
\left[\p_\theta + {1\over 2}\cot\theta - m \, (\sin\theta)^{-1} \,
\sigma^3 - i\sigma^2\right] \Psi_v =0\, .
\ee
This has the following non-singular solutions:
\ben \label{erew4}
\Psi_v &\propto& e^{i\phi/2}\,
\pmatrix{\cos(\theta/ 2) \cr -\sin(\theta/ 2)}
\quad \hbox{for} \quad m={1\over 2}\, , \nn
\Psi_v &\propto& e^{-i\phi/2}\,
\pmatrix{\sin(\theta/ 2) \cr \cos(\theta/ 2)}
\quad \hbox{for} \quad m=-{1\over 2}\, ,
\een
where the `constants' of proportionality could involve arbitrary
functions of $r$ and $v$.
Note that we have included in \refb{erew4} the
$\phi$ dependence of $\Psi_v$.
On the other hand the equation obtained by setting to zero the first
line of \refb{erew1} gives
\be \label{erew5}
\left[\p_r + {\psi'\over \psi}
-{1\over 2} \,   {\psi'\over \psi}\, \wh \Gamma^1\right]\Psi_v
= 0\, .
\ee
Now since $(\wh \Gamma^1)^2=1$, $\wh \Gamma^1$ 
has eigenvalues $\pm 1$.
Thus we can try to solve this equation separately in the sector
with $\wh\Gamma^1$ eigenvalue 1 and $\wh\Gamma^1$ eigenvalue
$-1$. The solutions are
\ben \label{erew6}
\Psi_v &=& \psi^{-3/2} \, \eta(v,\theta,\phi)
\quad \hbox{for} \quad
\wh\Gamma^1\, \eta = -\eta\, , \nn
\Psi_v &=& \psi^{-1/2} \, \eta(v,\theta,\phi)
\quad \hbox{for} \quad
\wh\Gamma^1 \eta = \eta\, ,
\een
where $\eta(v,\theta,\phi)$ is an SO(5,1) spinor and also an
$SO(5)$ spinor. The
$(\theta,\phi)$ dependence of $\eta(v,\theta,\phi)$  was
computed in \refb{erew4}
and the $v$ dependence is arbitrary except
for the periodicity requirement imposed by the period of the
coordinate $x^5$.
Both these solutions vanish as we approach the horizon although the
first solution vanishes more rapidly. Thus at this stage both would
appear to be acceptable solutions. However we shall see in \S\ref{s4}
that only the first solution preserves supersymmetry and hence only
these deformations will contribute to the index.
Furthermore we shall see in appendix \ref{sappc} that the second
solution is singular 
at the future horizon and hence
should not be counted as a true hair degree of freedom.
Thus $\Psi_v$
satisfies
\be \label{erew6.5}
\wh\Gamma^1 \Psi_v = -\Psi_v\, ,
\ee
and
the deformations associated with the gravitino
field take the  form:
\begin{equation}
  \label{eq:20pre}
  \Psi_v = \psi^{-3/2}(r)\, \eta(v,\theta,\phi), \qquad
  (\wt\Gamma^0 + \wt\Gamma^1)\, \eta(v,\theta,\phi) = 0\, ,
  \qquad \wh\Gamma^1 \eta(v,\theta,\phi)=-\eta(v,\theta,\phi)\, .
\end{equation}

Note that since $\psi\to 1$ as $r\to\infty$, the solution is
 $r$ independent at infinity and hence is not  normalizable. This
however can be rectified by making a local
supersymmetry transformation with a parameter that
approaches $-\int^v dv' \eta(v',\theta,\phi)$ as $r\to\infty$
and which vanishes sufficiently fast as we approach the
horizon. This sets the $r$ independent part of
the gravitino to zero at infinity but does not affect the mode
near the horizon.

Let us now count the number of independent functions
characterizing this deformation. To begin with $\Psi_v$ is an
$8\times 4=32$ dimensional complex spinor. But since $\Psi_v$
is a chiral spinor of $SO(5,1)$ only 16 of the 32 components
are independent. The two additional conditions listed in
\refb{eng}, \refb{erew6.5}
cut down the number of independent complex
components to 4. Finally we need to recall that the gravitino
field satisfies a symplectic Majorana
condition\cite{romans,9712176,9804166}. This gives altogether
four independent real functions of $v$ 
labelling the deformation, as expected.
It follows from \refb{erew4} that half of these deformations carry
$1/2$ unit of $\phi$ momentum and the other half carries
$-1/2$ unit of $\phi$ momentum.

Finally note that if we switch on the deformation \refb{eq:5}
then the extra terms containing at least one power of $T(v,\vec w)$
and one power of $\Psi_v$ will be of weight 3 and higher since the
deformation associated with $T(v,\vec w)$ is of weight 2. However
the gravitino equation \refb{eq:18} does not have any weight 3
component. Thus we conclude that the extra terms proportional
to \refb{eq:5} cannot affect the solution for the gravitino.

\sectiono{Four Dimensional Black Hole Hair} \label{s3}

We shall now consider the case of four dimensional black hole
obtained by placing the five dimensional black hole at the center
of the Taub-NUT space. For this we introduce the Taub-NUT metric
\be \label{ep7}
ds_{TN}^2 =  \left({4\over R_4^{2}}+{1\over r}\right)^{-1}
(dx^4 +\cos\theta d\phi)^2 + \left({4\over R_4^{2}}
+{1\over r}\right) \,
(dr^2 + r^2 d\theta^2 + r^2\, \sin^2\theta \, d\phi^2)\, ,
\ee
and replace the metric $dw^i dw^i$ in \refb{ep4}
by this Taub-NUT metric.
The full solution is given by\footnote{For $R_4^6< \wt J^2$  
the projection of this metric in the $x^4-x^5$ plane develops
a negative eigenvalue, giving rise to closed time-like curves.
We shall take $R_4^6>\wt J^2$ to avoid this situation.}
\ben \label{ep4tn}
ds^2 &=&  \psi^{-1}(r)
 \left[ du\, dv + \left(\psi(r)-1\right) dv^2
-2\wt\zeta\, dv
\right] +\psi(r)\, ds_{TN}^2\, ,  
\nonumber \\
e^{-2\Phi}&=&\lambda^{-2} \, , \nonumber \\
F^{(3)}&\equiv&
{1\over 6} \, F^{(3)}_{MNP} dx^M \wedge dx^N\wedge
dx^P \nonumber \\
&=& {r_0\over \lambda}\,
 \left[\left(\epsilon_3 +*_6 \epsilon_3\right)
+{1\over r_0} \left( 1 + { r_0   \over  r}\right)^{-1}
(dx^5+dt)\wedge d \wt\zeta\right] \, , \nonumber \\
\psi(r) &\equiv& \left( 1 + { r_0  \over r } \right)\, ,
\nonumber \\
\wt\zeta &\equiv&  -{\wJ \over 8} \left({1\over r}
+{4\over R_4^{2}}\right) \, (dx^4 +\cos\theta d\phi)\, , \nonumber \\
\eps_3 &\equiv&  \sin\theta\, dx^4 \wedge d\theta\wedge d\phi\, ,
\quad u\equiv x^5-t, \quad v \equiv x^5+t\, .
\een
It will be convenient to also introduce the coordinates
\be \label{eydef}
y^1 = r \, \sin\theta\, \cos\phi, \qquad y^2 = r \, \sin\theta\,
\sin\phi, \qquad y^3=r\, \cos\theta\, .
\ee
For $r>> R_4$ the asymptotic space-time locally 
has the form of
$K3\times S^1\times \wt S^1\times \RRR^{3,1}$, with
$x^4$ labelling the coordinate along the circle $\wt S^1$ and
$(y^1,y^2,y^3)$ labelling the space-like directions of $\RRR^{3,1}$.
$\wt S^1$ is non-trivially fibered over the
boundary $S^2$ of $\RRR^3$ reflecting that the
space is actually Taub-NUT.  
Even in this modified background
the field strength $F^{(3)}$   is self-dual, and hence
all the
anti-self-dual field strengths $H^s_{MNP}$'s
continue to vanish.\footnote{Note that the asymptotic
metric has a $dx^4 dt$ component. This can be removed by a
shift of the $x^4$ coordinate proportional to $t$ followed
by a rescaling of $t$.}

We can take the near horizon limit of \refb{ep4tn} using the same
coordinates introduced in \refb{ep8} and taking the $\beta\to 0$
limit. It is easy to see that in this limit the solution \refb{ep4tn}
reduces to \refb{ep9}. Thus \refb{ep4} and \refb{ep4tn} have the
same near horizon geometry\cite{0503217,0901.0359}.
We also show in appendix
\ref{sappc} that these solutions are non-singular
at the future horizon.

\subsection{Bosonic deformations representing transverse oscillation
of the black hole}  \label{s3.1}

We can generate deformations describing the oscillation of the
black hole in the three transverse non-compact direction as in
\S\ref{s2.1}. In particular we deform the metric to
\ben \label{emdef}
ds^2 &=&  \psi^{-1}(r)
 \left[ du\, dv + \left(\psi(r)-1 + \wt T(v,\vec y , x^4)\, \right) dv^2
-2 \wt\zeta\, dv
 \right]  +\psi(r)\, ds_{TN}^2\, .  
\een
Again the argument below \refb{elaplace} tells us that without
any modification of the scalar and the 3-form fields, \refb{emdef}
is guaranteed to be a solution to the equations of motion
if $\wt T(v,\vec y , x^4)$ is harmonic in the Taub-NUT space. Now one can
verify that acting on an $x^4$ independent configuration
the Laplacian
in the Taub-NUT space is proportional to $\vec \nabla_y^2$
\i.e.\ the laplacian in flat three dimensional
space labelled by the Cartesian coordinates $(y^1,y^2,y^3)$.
\be \label{efourdef}
\wt T(v, \vec y , x^4) \equiv \wt T(v, \vec y) = \vec g(v)\cdot \vec y\, ,
\qquad \int_0^{2\pi R_5} g_i(v) dv =0\, ,  
\ee
where $(g_1(v), g_2(v), g_3(v))$ are three arbitrary functions
subject to the restriction described above. 
These generate deformations representing transverse oscillation
of the black hole with finite amplitude. Again one can show that
even though the corresponding metric is not asymptotically flat,
one can make a coordinate transformation
\ben \label{ebmpvcoor}
&& v=v', \qquad \vec y = \vec y{~'}+\vec F, \qquad
u = u' - {8\over R_4^2}\dot F_i \, y^{\prime i} - {8\over R_4^2} 
\dot F_i \, F_i + {4\over R_4^2}\int^{{v'}}
\dot F_i(v'')\, \dot F_i(v'') dv'', \nn &&
{8\over R_4^2} \ddot F_i(v) \equiv g_i(v)\, ,  
\een
 to bring it to the
asymptotically flat form. Furthermore
to linear order in $\vec g(v)$ these deformations reduce to
those given in \cite{0901.0359} after a
coordinate transformation.  Finally to check that the deformations
represent hair modes we note that in the coordinate system \refb{ep8}
they scale as $\beta^2$ and hence vanish as $\beta\to 0$.
We also show in appendix
\ref{sappc} that unlike the deformations described
in \S\ref{s2.1}, these solutions are non-singular
at the future horizon. The main difference between the deformations
described in \S\ref{s2.1} and those described here is due to the fact
that the former were proportional to $\vec f\cdot \vec w$ which vanish
as $\sqrt r$ as $r\to 0$, whereas the latter, being proportional to
$\vec g\cdot \vec y$, vanish as $r$ as $r\to 0$.

\subsection{Bosonic deformation representing the oscillation of the
2-form fields} \label{s3.2}

Taub-NUT space has a self-dual harmonic form $\omega_{TN}$
given by
\be \label{eom1}
\omega_{TN} = -
{r \over 4r+ R_4^{ 2} } \,
\sin\theta d\theta\wedge d\phi
+{R_4^{ 2} \over (4r+ R_4^{ 2})^2} \,
 dr\wedge
(dx^4 + \cos\theta d\phi)\, .
\ee
Now as was discussed at the beginning of \S\ref{s2},
in type IIB string theory on $K3$
we have 21 2-form fields with anti-self-dual field strength,
collectively denoted as $H^s_{MNP}$ ($1\le s\le 21$).
We now switch on a deformation of the form\cite{0901.0359}
\be \label{enewf}
\delta (ds^2) =  \psi^{-1}(r)
\left(\wt T(v,\vec y)+ \wt S(v,\vec y , x^4)\, \right) dv^2, \qquad
 \delta H^s = h^s(v) \ dv\wedge\omega_{TN}\, ,
\ee where $h^s(v)$ are arbitrary functions, $\wt T(v,\vec y)$ is
given in \refb{efourdef}, and $\wt S$ is quadratic in $h^s$ and will
be determined below. We can verify, first of all, that the
deformations $\delta H^s$ given in \refb{enewf} are closed and
satisfy the requirement of anti-self-duality even in the presence of
the metric deformation parametrized by $\wt T + \wt S$. Now it was
shown in \cite{0901.0359} that to linearized order the deformation
\refb{enewf} satisfy the equations of motion without any need to
modify the background metric or the scalars due to the relation \be
\label{ereason} F^{(3)}_{MPQ} \delta H^{sPQ}_N = 0\, , \ee so that
the cross terms between the background 3-form field and the
deformations do not produce a source for the metric and scalars.
Thus we only need to analyze the contribution to the equations of
motion from higher order terms. It follows from the arguments below
\refb{elaplace} and that fact that under \refb{elaplace} the
deformations  $\delta H^s$ and $\delta (ds^2)$ have 
weights one and
two respectively that the only non-trivial equation that we need to
check at quadratic and higher order in the deformation is the $vv$
component of the metric equation. Furthermore the equation should
involve at most linear terms in $\wt T$ and $\wt S$ without any
power of $h^s$ or two powers of $h^s$ without any factor of $\wt T$
or $\wt S$. Expressing the metric equation as $R_{MN}\propto T_{MN}$
and the fact that the contribution to $T_{MN}$ from the $H^s$ fields
is proportional to $H^s_{MPQ} H^{sPQ}_N$ at the quadratic order, we
get\footnote{It is a little easier to use the metric equation of the
form $R^u_v\propto T^u_v$. In this case $\delta R^u_v= \psi^{-1}\,
\nabla_\perp^2 (\wt T+\wt S)=\psi^{-1}\nabla_\perp^2\wt S$ and
$\delta T^u_v\propto \psi^{-1} \, (4r + R_4^2)^{-4}$, leading to
\refb{eq:15}.}
\begin{equation}
\label{eq:15} \nabla_\perp^{2} \wt S(v,\vec y, x^4) = C(v) \, R_4^2\, (4r
+ R_4^2)^{-4}\ , \qquad C(v) \equiv 8\, h^s(v)\, h^s(v)\, ,
\end{equation}
where  $\nabla_\perp^2$ denotes the Laplacian in the Taub-NUT space.
In arriving at \refb{eq:15} we have used the fact that
$\nabla_\perp^2 \wt T(v,\vec y)=0$. Since any solution to the source
free equation can be absorbed into $\wt T$ we only need to look for
a particular solution. The following solution describes a
normalizable deformation of the metric living outside the horizon:
\be \label{enorm}
\wt S(v,\vec y, x^4) = \frac{C(v) r}{2R_4^2(4r+R_4^2)}
\ee
The $\wt S(v)$ given in \refb{enorm} does not vanish at infinity
but this can be easily repaired by an appropriate reparametrization
which takes the form $u\to u-{1\over 8 R_4^2}\int^v C(v') dv'$ 
as $r\to\infty$ and $u\to u$ as $r\to 0$.
Finally to check that these deformations represent hair modes we
note that in the coordinate system \refb{ep8}, $\delta H^s$ given in
\refb{enewf} scales as $\beta$ and hence vanishes as $\beta\to 0$.

This shows that we have a family of finite deformations, labelled
by the 24 functions $(\vec g(v), \{h^s(v)\})$, of the original four
dimensional black hole solution. Furthermore these deformations
are supported outside the horizon and do not affect the
horizon geometry.
We also show in appendix
\ref{sappc} that these solutions are non-singular
at the future horizon.

\subsection{Fermionic deformations} \label{s3.3}

Construction of the left-moving fermionic deformations proceeds
in the same way as in \S\ref{s2.2}. The analysis up to
\refb{egr3} is more or less identical except that we now have
different expressions for the
simplified form of vielbeins and the
spin connections after dropping terms of weight $>0$. Thus
equations
\refb{esimpleviel},
\refb{espinsimple} get replaced by:
\ben \label{efourviel}
&& e^0 =  {1\over 2}  (dv-\psi^{-1}du),
\qquad e^1 = {1\over 2}   (dv+
\psi^{-1}du),
\nonumber \\
&& e^2 = \psi^{1/2}r^{1/2}\chi^{-1/2}
(dx^4 + \cos\theta d\phi), \qquad
 e^3 = \psi^{1/2}r^{-1/2}\chi^{1/2}dr,
\nonumber \\
&& e^4 = \psi^{1/2}r^{1/2}\chi^{1/2}d\theta, \qquad e^5=
\psi^{1/2}r^{1/2}\chi^{1/2}\sin\theta\, d\phi\, ,
\een
\be \label{edefchi}
\chi \equiv \left(1  +{4r\over R_4^2}\right)\, ,
\ee
\ben \label{espinfour}
&& \omega_r^{01} = -{1\over 2}\, {\psi'\over \psi}\, ,
\quad \omega_{x^4}^{23} = {1\over 2}\, {(r\psi\chi^{-1})'\over \psi}\, ,
\quad \omega_{x^4}^{45}={1\over 2}\chi^{-2}, \quad
\omega_\phi^{23} = {1\over 2}\, {(r\psi\chi^{-1})'\over \psi}\,
\cos\theta,\nonumber \\
&&
\omega_\phi^{24} = - {1\over 2}\, \chi^{-1}\,
\sin\theta,  \quad
\omega_\phi^{35} =- {1\over 2}\, {(r\psi\chi)'\over \psi\chi}
\, \sin\theta,
\quad \omega_\phi^{45}= - {1\over 2}\, (2-\chi^{-2})\,
\cos\theta\, , \nn &&
\omega_\theta^{25} = {1\over 2}\, \chi^{-1}, \quad
\omega_\theta^{34} = -{1\over 2} \, {(r\psi\chi)'\over \psi\chi}\, .
\een

Substituting these into the gravitino
 equation and
 assuming that these modes
have no  $x^4$ dependence
 we get the following form.
\begin{eqnarray}
  \label{eq:1}
&&  \sqrt{\frac{r}{\psi\chi}}\tilde\Gamma^3
\left[ \partial_r + \frac{\psi'}{\psi}
-\frac{\psi'}{2\psi}\wh\Gamma^1 + \frac{3}{4r}
+\frac{\chi'}{4\chi} + \frac{1}{4r\chi}
\right]\Psi_v\nonumber \\
&&+ \frac{1}{\sqrt{r\psi\chi}}\tilde\Gamma^4
\left[\partial_\theta + \frac{\cos\theta}
{2\sin\theta}\right]\Psi_v +
\frac{1}{\sqrt{r\psi\chi}\sin\theta}\tilde
\Gamma^5\partial_\phi
\Psi_v = 0\, .
\end{eqnarray}
Using the fact that
\be \label{echirela}
{\chi'\over 4\, \chi} + {1\over 4\, r\, \chi} = {1\over 4\, r}\, ,
\ee
we can see that the $R_4$ dependence drops out and hence
eq.\refb{eq:1} is  identical to the corresponding equations for the
BMPV black hole in flat transverse space.
 The gravitino modes
are therefore unaffected by the Taub-NUT space!
Finally, as in the case of BMPV black holes in flat space,
the gravitino
zero modes in this case are also
non-singular
at the future horizon.

When we switch on the deformations described in \S\ref{s3.2},
$H^{sMNP}$ no longer vanish and we need to examine the
second equation of \refb{eq:18}. However since the deformation
given in \refb{enewf} has weight $\ge 1$, it can only contribute to
equations with weight $\ge 1$. The right hand side of the second
equations in \refb{eq:18} however has weight 0. Thus we conclude
that this equation is not affected by the deformations given
in \refb{enewf}.

\subsection{Bosonic deformation representing relative oscillation
between the BMPV black hole and KK monopole} \label{s3.4}

It was argued in \cite{0901.0359} that the BMPV black hole
in Taub-NUT space contains another set of hair degrees of freedom
which represent the left-moving
oscillations of the BMPV black hole relative to the
Taub-NUT background. In the limit when the Taub-NUT radius goes
to infinity 
these modes coincide with the transverse
left-moving oscillation modes
of the BMPV black hole in flat space-time, constructed
in \S\ref{s2.1}. Also since near the origin
the Taub-NUT metric looks like flat metric, we expect that near the
horizon these modes will have the same behaviour as the
transverse oscillation modes of the BMPV black hole in flat space-time.

We have not tried to construct these modes explicitly for finite $R_4$
since, as these modes have identical near horizon behaviour as those
of \S\ref{s2.1}, they will have a curvature singularity at the
future horizon. Thus we shall not count these modes
among the hair degrees of freedom of the BMPV black hole in
Taub-NUT space.

\sectiono{Supersymmetry of the Deformed Configuration}
\label{s4}

In order to study the supersymmetry of the deformed background
we need to examine the equations which set to zero the
supersymmetry variation of all the fields. Since we shall
always be working in a background where scalar fields are
constants and the spin 1/2 fields $\chi^{\alpha r}$ are zero, we shall
write down the equations in this background.
The equations obtained by setting to zero the supersymmetry
variation of the metric,
the gravitino, the 3-form field strength $\bar H^k_{MNP}$
and the spin 1/2 field $\chi^{\alpha r}$
take the form\cite{romans,9712176,9804166}:
\ben \label{esusytrs}
&& \eta_{AB}\, \bar\eps \,  \wt\Gamma^A \, \Psi_M \, e_N^{~B}
+ \eta_{AB}\, e_M^{~A}\, \bar\eps \,
\wt\Gamma^B \, \Psi_N= 0
\, , \nn
&& D_M\eps -{1\over 4}\bar H^i_{MNP} \, \Gamma^{NP}
\, \wh\Gamma^i\eps=0\, , \nn
&&\p_{[P}
\left(\bar\eps \, \Gamma_{M} \, \wh\Gamma^i \Psi_{N]}
\right)=0\, , \nn
&&\Gamma^{MNP} H^s_{MNP} \eps= 0\, ,
\een
where $\eps$ is the supersymmetry transformation parameter satisfying
\be \label{eepscons}
(\wt\Gamma_{012345} + 1) \eps =0, \qquad \bar \eps
= \eps^T \, C \, \Omega\, ,
\ee
$C$ and $\Omega$ being the $SO(5,1)$ and $SO(5)$ charge
conjugation matrices satisfying
\be \label{ecconj}
(C\wt\Gamma^A)^T = - C\wt\Gamma^A, \qquad
(\Omega \wh\Gamma^i)^T = - \Omega \wh\Gamma^i, \qquad
\Omega^T = -\Omega\, .
\ee
The equations obtained by setting to zero the variations of the
fields strengths $H^s_{MNP}$ and the scalar fields are
automatically satisfied in this restricted class of backgrounds.
Note that we have suppressed the SO(5,1) and SO(5) spinor
indices in eqs.\refb{esusytrs}.

Now suppose $\eps_{(0)}$ is a Killing spinor of the original
background which could be either the BMPV black hole in flat
transverse space or BMPV black hole in Taub-NUT space. The explicit
form of $\eps_{(0)}$ has been given in appendix \ref{sec:appendix},
but we only need to use the fact that it satisfies the projection
conditions
\be \label{eproj} (\wt\Gamma^0+\wt\Gamma^1)\, \eps_{(0)}=0
\quad \to \quad \Gamma^v \, \eps_{(0)} = 0\, ,
\ee
and
\be
\label{esecondproj} \wh\Gamma^1 \eps_{(0)} =\eps_{(0)}\, .
\ee
Since $\eps_{(0)}$ does not have any space-time index we can assign to it
weight zero in the convention described below \refb{elaplace}.
Furthermore due to \refb{eproj} we cannot reduce the weight of any
expression containing $\eps_{(0)}$ by acting on it by $\Gamma^v$. We
shall now see that this guarantees that $\eps_{(0)}$ automatically
satisfies \refb{esusytrs} even in the presence of the deformations.

We begin with the metric deformations. Since these deformation are
proportional to $dv^2$, they carry weight 2. Thus any term involving
these deformations will generate terms of weight $\ge 2$.
Examining \refb{esusytrs} we see that the only term that carries
weight $\ge 2$ is the weight 2 term obtained by choosing $M=N=v$ in
the first equation. But this is  linear in gravitino deformation
$\Psi_v$ which
already carries weight 1. Since a term linear in $\Psi_v$ and
also in $\delta G_{vv}$ has weight 3, we see that
$\delta G_{vv}$ cannot affect
the first  equation of \refb{esusytrs}. Thus  we
conclude that the metric deformations considered here are invariant
under $\eps_{(0)}$.

Next we turn to the deformations involving three form field strengths
$H^r_{MNP}$ as described in
\S\ref{s3.2}.
The only equation in \refb{esusytrs} which involves
$H^r_{MNP}$ is the last equation. However from \refb{enewf}
we see that $H^r_{MNP} \Gamma^{MNP}$ is proportional
to $\Gamma^v$ and hence the last term in \refb{esusytrs}
vanishes
identically due to \refb{eproj}.
On the other hand   the arguement given in the previous
paragraph shows that the induced metric \refb{enorm} does not
affect the Killing spinor equation.
Thus these deformations also
do not destroy the supersymmetry of the background.

Finally we turn to the fermionic deformations involving $\Psi_v$.
Since this has weight 1, it can only affect terms in the equation
with weight $\ge 1$. The relevant equations are the first and third
equation of \refb{esusytrs}.   
In the first equation we can choose $MN$
to be either $vv$ or $vw^i$, and in the third equation we need to
choose $PMN$ to be $w^iw^jv$. Now the first equation involves
terms of the form $\bar\eps_{(0)} \wt\Gamma^A \Psi_v$. Since
$\eps_{(0)}$ and
$\Psi_v$ satisfy opposite $\wh\Gamma^1$ projection (see
eqs.\refb{erew6.5} and \refb{esecondproj}) these terms vanish.
Thus we only need to examine the left hand side of the third
equation. It follows from \refb{egravchiral}, \refb{eng} and
\refb{eepscons}, \refb{eproj} that $\eps_{(0)}$ and $\Psi_v$ satisfiy
the same  $SO(4)$ projection rules:
\be \label{eso4proj}
\wt\Gamma^{2345}\Psi_v=\Psi_v, \qquad
\wt\Gamma^{2345}\eps_{(0)} = \eps_{(0)}\, .
\ee
As a result $\bar\eps_{(0)} \wt\Gamma^i
\wh\Gamma^k \Psi_v$ vanishes for $i=2,3,4,5$.
This in turn shows that the left hand side of the third term also
vanishes.

\sectiono{Partition Function After Hair Removal} \label{s5}

In this section we shall briefly analyze the partition functions of the
five and the four dimensional black hole entropy after hair removal.
The analysis will be similar to that in \cite{0901.0359} except that
we shall now take into account the fact that the plane waves describing
the transverse oscillation of the BMPV black hole have curvature
singularities at the future horizon and hence should not be counted as
part of the hair degrees of freedom. Also for simplicity we shall ignore
the contribution to the degeneracies from small black hole core dressed
by hair, -- a detailed discussion on this can be found in
\cite{0901.0359}. The net effect of this is to remove from the
final partition functions \refb{ese2}, \refb{eadd2}
the contribution from
the half-BPS states. Even at the intermediate stages of the
analysis these contributions are exponentially suppressed
compared to the leading contribution.

We consider the case where there is a single D5-brane, and introduce
the variables $(\rho,\sigma,v)$ as conjugates to the D1-brane charge
along $S^1$, momentum along $S^1$ and the momentum along $x^4$.
The index
is related to the partition function $Z$ by Fourier
transform.\footnote{For a precise definition of what
index and partition function we are 
computing see ref.\cite{0901.0359}.}  
Then the microscopic partition function of the five
dimensional system is computed by multiplying the partition
function associated with the oscillations of the D1-branes relative to
the D5-brane\cite{9608096}
and the center of mass oscillation of the
combined system. The result
is\cite{0901.0359}\footnote{The terms in the expansion whose
$\rho$ dependence is of the form $e^{-2\pi i\rho}$ represent
contribution from half BPS states. We can remove this contribution by an
appropriate subtraction as was done in \cite{0901.0359}, but for
simplicity we shall ignore this complication.}
\ben\label{ep2}
Z_{5D}(\rho,\sigma,v) &=&
e^{-2\pi i
\rho-2\pi i \sigma
 }\prod_{k,l,j\in \zzz
\atop k\ge 1, l\ge 0}
\left( 1 - e^{2\pi i ( \sigma k   +  \rho l + v j)}
\right)^{-c(4lk - j^2)} \nonumber \\
&& \times \left\{\prod_{l\ge 1} (1-e^{2\pi i(l\rho+v)})^{-2} \,
(1-e^{2\pi i(l\rho-v)})^{-2} \, (1-e^{2\pi il\rho})^4 \right\}\,
(-1)\, (e^{\pi i v}-
e^{-\pi i v})^2\, . \nonumber \\
\een
The hair of the five dimensional black hole contains a set of 12
gravitino zero modes. Their  
quantum numbers can be easily read
out from the quantum numbers associated
with the broken supersymmetries.
Four of these zero modes do not carry any
$x^4$ momentum (which in five dimensions is a particular component
of the angular momentum) -- they are used in soaking up the fermion
zero modes in the computation of the helicity trace\cite{0901.0359}.
The rest carry $x^4$ momentum $\pm 1/2$ and gives a contribution to
the partition function of the form
$(e^{\pi i v}-e^{-\pi i v})^4$\cite{0901.0359}.
Finally there are 4 left-moving gravitino modes carrying no $x^4$
momentum as described in \S\ref{s2.2}. They give a contribution
of $\prod_{l\ge 1} \, (1-e^{2\pi il\rho})^4$.
Combining these
two contributions we get
\be \label{ese1}
Z^{hair}_{5D}(\rho,\sigma,v)
=  (e^{\pi i v}-e^{-\pi i v})^4\,
\prod_{l\ge 1}  \, (1-e^{2\pi il\rho})^4 \, .
\ee
Hence the partition function associated with the horizon is given by
\ben \label{ese2}
Z_{5D}^{hor}(\rho,\sigma,v) &= & Z_{5D}
/Z_{5D}^{hair} \nonumber \\
&=& -e^{-2\pi i
\rho-2\pi i \sigma
 }\, (e^{\pi i v}-e^{-\pi i v})^{-2}\,
 \prod_{k,l,j\in \zzz
\atop k\ge 1, l\ge 0}
\left( 1 - e^{2\pi i ( \sigma k   +  \rho l + v j)}
\right)^{-c(4lk - j^2)} \nn
&& \left\{\prod_{l\ge 1} (1-e^{2\pi i(l\rho+v)})^{-2} \,
(1-e^{2\pi i(l\rho-v)})^{-2}  \right\}\, .
\een

Let us now repeat the analysis for the BMPV black hole in transverse
Taub-NUT space. The microscopic partition function is given
by\cite{9607026,0412287,0505094,0506249,0605210}
\be\label{ep2a}
Z_{4D}(\rho,\sigma,v) =
-e^{-2\pi i
\rho -2\pi i\sigma -2\pi i v
 }\prod_{k,l,j\in \zzz
\atop k,l\ge 0, j<0 \, for\, k=l=0}
\left( 1 - e^{2\pi i ( \sigma k   +  \rho l + v j)}
\right)^{-c(4lk - j^2)}\, .
\ee
In this case the hair modes include 12 fermion zero modes all of
which are used in saturating the helicity factors inserted
into the helicity trace. Besides these there are 21 left-moving
bosonic modes associated with the 2-form deformations and
3 left-moving bosonic modes associated with the transverse oscillation
of the black hole, all of which are neutral under the $x^4$
translation. Finally there are four left-moving gravitini modes,
also neutral under $x^4$. These four fermionic modes cancel the
contribution from four of the bosonic modes and we are left with the
contribution:
\be \label{eh1}
Z_{4D}^{hair}(\rho,\sigma,v) = \prod_{l=1}^\infty
\left(1 - e^{2\pi i l \rho}\right)^{-20}
\, .
\ee
Thus we get
\ben \label{eadd2}
Z_{4D}^{hor}(\rho,\sigma,v) &=& Z_{4D}
/Z_{4D}^{hair}\nonumber \\
&=& -e^{-2\pi i
\rho-2\pi i \sigma
 }\, (e^{\pi i v}-e^{-\pi i v})^{-2}\,
 \prod_{k,l,j\in \zzz
\atop k\ge 1, l\ge 0}
\left( 1 - e^{2\pi i ( \sigma k   +  \rho l + v j)}
\right)^{-c(4lk - j^2)} \nn
&& \left\{\prod_{l\ge 1} (1-e^{2\pi i(l\rho+v)})^{-2} \,
(1-e^{2\pi i(l\rho-v)})^{-2}  \right\}\, ,
\een
where is the last step we have used $c(-1)=2$, $c(0)=20$.
Comparing \refb{eadd2} with \refb{ese2} we see that
\be \label{eequal}
Z_{5D}^{hor}(\rho,\sigma,v) = Z_{4D}^{hor}(\rho,\sigma,v)\, .
\ee


\bigskip

{\bf Acknowledgement:} We would like to thank Nabamita Banerjee,
Borundev Chowdhury, Justin David, Suresh Govindarajan, Ipsita Mandal,
Samir Mathur, Ashish Saxena and Nemani Suryanarayana for useful
discussions.  DPJ and YS would like to thank CHEP, Bangalore for warm
hospitality during the course of this work.  DPJ would also like to
thank IIT, Madras and IMSc, Chennai for warm hospitality.  
A.S. would like to thank GGI, Florence and University of Roma,
Tor Vergata for warm
hospitality during the course of this work.  
This work
was supported by the project 11-R\& D-HRI-5.02-0304 and the J.C.Bose
fellowship of the Department of Science and Technology.

\appendix

\sectiono{Killing Spinors}
\label{sec:appendix}

The Killing spinor equation in the BMPV black hole and BMPV black hole
in the Taub-NUT space, obtained by setting $\delta\Psi_M^\alpha=0$,
 is
\begin{equation}
  \label{eq:6}
  D_M\epsilon - \frac{1}{4}\bar H^i_{MNP}
  \Gamma^{NP}\wh\Gamma^i\epsilon =
  0\, ,
\end{equation}
where $\bar H^i_{MNP}$ for $1\leq i \leq 5$ are self-dual field
strengths of 2-form fields in six dimensions.
The equations obtained by setting $\delta\chi^{\alpha r}=0$ involve
anti-self-dual components of the 3-form field strength and are
automatically satisfied in this background.
The three form field
strength $F^{(3)}$ appearing in the BMPV black hole in the flat
transverse space and BMPV in the Taub-NUT space is expressed in terms
of $\bar H^i$ as
\begin{equation}
  \label{eq:11}
  F^{(3)}_{MNP} = 2\, \lambda^{-1}\, \bar H^1_{MNP}\, .  
\end{equation}
As described below \refb{elaplace}, we can decompose the background
field configuration as sum of different components carrying different
weights.  The weight of any term appearing in the left hand side of
the Killing spinor equation is greater than or equal to the sum of the
weights of the various field components which enter that term.  This
is due to the fact that the only way to reduce the weight of a given
combination of fields is to contract one of the covariant $v$ index
with a $\Gamma^v$ but we prevent this from happening by demanding that
\begin{equation}
  \label{eq:12}
  \Gamma^v\epsilon = 0\,  \quad \to \quad \wt\Gamma^0\wt\Gamma^1
  \eps=\eps\, .
\end{equation}
Now since the $v$ component of the Killing spinor equation has weight
1, the $u$ component has weight $-1$ and the other components have
weight zero we conclude that for $M=v$ the left hand side of
eq.\refb{eq:6} can receive contribution only from terms of weight zero
or one in the field configuration, for $M=u$ the equation must be
identically satisfied and for $M\ne u, v$ only terms of weight 0 can
contribute.  In particular since terms in $F^{(3)}$ which are
independent of $\tilde J$ are of weight zero and the $\tilde J$
dependent terms have weight one, we see that $\tilde J$ dependent
pieces do not contribute to the Killing spinor equation for $M\ne v$.
However $\tilde J$ dependent terms could potentially contribute to the
$v$ component of the Killing spinor equation.

Let us first look at the $v$ component of the Killing spinor equation.
Requiring the Killing spinor to be $v$ independent we find that for
the BMPV black hole in flat transverse space
this equation takes the form
\be \label{evcomp}
\left[{\psi'\over 8\psi^2}\Gamma^{ru} - {1\over 8} \p_{w^i}
  (\psi^{-1}\chi_j) \Gamma^{ij} - {\psi'\over
    8\psi^2}\Gamma^{ru}\wh\Gamma^1 + {1\over 8} \p_{w^i}
  (\psi^{-1}\chi_j) \Gamma^{ij}\wh\Gamma^1\right] \eps = 0\, .
\ee
This can be satisfied by choosing
\be \label{evsoln}
\wh\Gamma^1\eps =\eps\, .
\ee
For BMPV black hole in Taub-NUT space $\chi_i$ in eq.\refb{evcomp} is
replaced by $-2\wt\zeta_i$, but \refb{evsoln} still provides a solution
to this equation.

Next we examine the $r$, $\theta$, $\phi$ and $x^4$
components of the Killing spinor equation.
Using \refb{evsoln}
and the spin connection components given in (\ref{espinsimple}) and
(\ref{espinfour}) for the BMPV black hole in the flat space and the
BMPV black hole in TN space respectively, we get
the following form of
these equations for both black holes:
\begin{eqnarray}
  \label{eq:13}
    \left(\partial_r + \frac{\psi'}{2\psi}\right)\epsilon &=&0\, , \\
  \left( \partial_\theta
    -\frac{1}{2}\wt\Gamma^{34}\right)\epsilon &=&0\, , \\
  \partial_{x^4}\epsilon &=&0\, , \\
  \left(\partial_\phi  -{1\over 2}\sin\theta \, \wt\Gamma^{35}
  -{1\over 2} \cos\theta \, \wt\Gamma^{45}\right)
  \epsilon &=& 0\, .
\end{eqnarray}
Using the gamma matrix representations given in \refb{erew2}
we find the following solutions:
\be \label{ekillsoln}
\eps = \psi(r)^{-1/2}\, e^{i\phi/2} \, \pmatrix{\cos(\theta/2)\cr
-\sin(\theta/2)}, \qquad \eps =
\psi(r)^{-1/2}\, e^{-i\phi/2} \, \pmatrix{\sin(\theta/2)\cr
\cos(\theta/2)}\, .
\ee
To count the number of independent Killing spinors we note that to
begin with $\eps$ is an $8\times 4=32$ dimensional complex spinor.
The chirality projection condition \refb{eepscons} and the two dynamical
constraints given in \refb{eq:12}, \refb{evsoln}
reduce this number to 4 complex
parameters. Finally a reality condition (symplectic Majorana) reduces
the number to 4 real parameters.

Note that the Killing spinor (\ref{ekillsoln}) is independent of
whether we consider BMPV black hole on flat transverse space or
Taub-NUT space.  This behaviour can be explained using the following
reasons. The Taub-NUT space has $SU(2)$ holonomy, which by convention
is identified with $SU(2)_L$ subgroup of its $SO(4)$ tangent space
symmetry.  Fermions in the Taub-NUT space transform as $(2,1)+(1,2)$
under $SO(4) = SU(2)_L\times SU(2)_R$.  Thus half of the fermions are
neutral under $SU(2)_L$ and hence behave as free fermions as far as
the Taub-NUT space is concerned.  In our six dimensional space, the
$SO(1,1)$ chirality is correlated with the $SO(4)$ chirality in the
following manner: $SU(2)_L$ singlets are left moving with respect to
$SO(1,1)$ and $SU(2)_R$ singlets are right moving.  Since Killing
spinors corresponds to unbroken supersymmetry, which in our convention
are left moving spinors of $SO(1,1)$, they are singlet of the Taub-NUT
holonomy group $SU(2)_L$.  As a result the Killing spinors are
unaffected when we replace flat space by the Taub-NUT space.

\sectiono{Black Hole Metric in Non-singular   
Coordinate System} \label{sappb}

In this appendix, following \cite{9605224,9606113}
we write the black hole metric in coordinates in which it
is regular and analytic at the future
horizon.  This coordinate system will
then be used in appendix
\ref{sappc} to analyse regularity of the modes at the location of the
horizon.    For simplicity we will be working
with $\tilde J=0$ solution. Later we shall briefly discuss the
extension to the $\wt J\ne 0$ case.

The original metric given in \refb{ep4} may be expressed as
\begin{equation} \label{eoror}
ds^2 = \psi^{-1}(dudv + K dv^2 ) + \psi\left(r^{-1}
dr^2  + 4\, r \, d\Omega_{3}^2\right)\, , \quad
d\Omega_3^2\equiv {1\over 4} \left((dx^4+\cos\theta d\phi)^2
+d\theta^2 +\sin^2\theta d\phi^2\right)\, ,
\end{equation}
where,
\begin{equation}
\psi = 1 + \frac{r_{0}}{r} , \qquad K = \psi -1 \, .
\end{equation}
Following \cite{9605224} we will
now do the following coordinate transformation:
\be
\label{newcoord} V= -\sqrt{r_0}\exp(-\frac{v}{\sqrt{r_0}}) ,
\qquad W=
\frac{1}{R}\exp(\frac{v}{2\sqrt{r_0}})  , \qquad
U = u + \frac{R^2}{2\sqrt{r_0}} + 2v\, ,
\ee
\be \label{edefr}
R\equiv 2\sqrt{r_0\left(1+ \frac{r_{0}}{r}\right)} \, .
\ee
Note that the region outside the horizon has $V<0$.
In these new coordinates the metric becomes
\begin{equation} \label{ebeco}   
ds^2  = 4\, r_0\, \left[W^2 dUdV   
+ dV^2\left\{\frac{3\sqrt{r_0} W^2}{V} -
\frac{(1-Z^{-3})}{4V^2} \right\} - dVdW \left(\frac{1-Z^{-3}}{VW}
\right) +
\frac{dW^2}{W^2Z^3} + \frac{d\Omega_{3}^2}{Z}\right] \, ,
\end{equation}
where
\be \label{ezdef}
Z\equiv 1 + 4\sqrt{r_0} VW^2.
\ee
 To see that metric is regular at $V=0$,
 we expand $Z$ in \refb{ebeco}  
to get
\begin{eqnarray} \label{eregm}
ds^2  &=&  4\, r_0\, \bigg[
W^2 dUdV + dV^2 r_0 W^4 Z^{-3} (24+
128\sqrt{r_0}V W^2 + 192 r_0 V^2 W^4)
 \nonumber \\
&& - dVdW  \, 4\sqrt{r_0} W}{Z^{-3}(3+ 12\sqrt{r_0}V W^2
+ 16r_0 V^2
W^4)
  +  W^{-2}Z^{-3}dW^2 + Z^{-1}d\Omega_{3}^2\bigg]\, . \nonumber \\
\end{eqnarray}
It is now easy to see
that the metric is regular at the future horizon
$V=0$. In fact the
metric components are polynomials in $V$ and therefore they are
analytic functions of $V$.
Thus all derivatives of the metric components, and hence the
Riemann tensor, remain finite at the horizon for
finite $W$.\footnote{For $V\to 0$ the metric reduces to that of
$AdS_3\times S^3$ locally\cite{9605224}.}

We can also write down the three form field strength in terms
of new coordinates.  For $\tilde J=0$ we get
\begin{equation}
  \label{eq:24}
  F^{(3)} = {r_0\over \lambda}\, \left[ \sin\theta\,
dx^4\wedge d\theta\wedge d\phi + 4WdW\wedge dV\wedge dU\right]\,  .
\end{equation}
In the near horizon limit, $F^{(3)}$ is well behaved and independent
of $V$.

Next we now look at the behaviour of
the $\wt J=0$ black hole in the Taub-NUT
space. The $G_{uv}$ component is the same as in the case of
the $\wt J=0$
black hole in flat transverse space. The difference comes in the
components involving $x^4$, $\theta$, $\phi$ and $r$ coordinates.
Most of these differences vanish near the horizon both in the
original coordinate system and in
the new coordinate system and hence
do not spoil the regularity property of the metric in the new
coordinates. The only additional term that apparently diverges 
at the
horizon is the contribution
\begin{equation}
  \label{eq:36}
  \delta (ds^2) = \frac{4\psi}{R_4^2}\, dr^2\, ,
\end{equation}
which in the new coordinate system becomes
\begin{equation}
  \label{eq:37}
  \delta (ds^2) = \frac{(4r_0)^3}{R_4^2 Z^4} \left( -\frac{W^2\,
   dV^2}{4
      \sqrt{r_0} V} -{V\, dW^2\over \sqrt{r_0}}
       - \frac{WdWdV}{\sqrt{r_0}}\right)\, .
\end{equation}
Among these, only the first term is singular at the horizon. This
singular contribution, however, can be removed by a shift in the
 $U$
coordinate of the form
$U\to U +4 r_0^{3/2} R_4^{-2}\ln(-V/\sqrt{r_0})$.
We can combine this shift with the one given in
\refb{newcoord} to write
\begin{equation}
  \label{eq:39}
  U = u + \frac{R^2}{2\sqrt{r_0}} + 2v + \frac{(4r_0)}{R_4^2}v\, .
\end{equation}
It is also easy to check that  $F^{(3)}$ has the same
form as \refb{eq:24} and hence is non-singular at the horizon.

For completeness we shall now briefly discuss the effect of
switching on the parameter $\wt J$ in the original solution,
labelling the angular momentum carried by the black hole. In this
case the metric has the form:
\ben \label{ebm1}
ds^2 &=& \psi^{-1}\left(dudv + K dv^2 + {\wt J\over 4r}
\, (dx^4+\cos\theta d\phi)
\, dv\right) \nn
&& + \psi\left[{dr^2\over r} + r
\left\{(dx^4+\cos\theta d\phi)^2 + d\theta^2 + \sin^2\theta d\phi^2
\right\}\right]\, .
\een
In order to introduce coordinates in which the metric at the future
horizon is non-singular, we first shift
$x^4 \to x^4 - {\wt J\over 8r_0^2}v$ so that the cross term between
$dv$ and $(dx^4+\cos\theta d\phi)$ 
has a zero at $r=0$.\footnote{Note that the periodic  
identification
$x^5\equiv x^5+2\pi R_5$ takes the form $(x^4,x^5)\equiv
(x^4+{\wt J\over 8 r_0^2} 2\pi R_5, x^5+2\pi R_5)$ in the
new coordinate system. This however does not affect our analysis.}
This brings the metric to the form:
\ben \label{ebm2}
ds^2 &=& \psi^{-1}\left[dudv + \left(K +{\wt J^2 \over 64 r_0^4}
r \psi^2 -{\wt J^2 \over 32 r_0^2 \, r}
\right) dv^2 + \left({\wt J\over 4r}
- {\wt J r \over 4 r_0^2}\psi^2\right)
\, (dx^4+\cos\theta d\phi)
\, dv\right] \nn
&& + \psi\left[{dr^2\over r} + r
\left\{(dx^4+\cos\theta d\phi)^2 + d\theta^2 + \sin^2\theta d\phi^2
\right\}\right]\, .
\een
In the next step we carry out a rescaling
$u\to (1-{\wt J^2\over 64r_0^3})^{1/2}
u$, $v\to v/(1-{\wt J^2\over 64r_0^3})^{1/2}$
so that the coefficient of the $dv^2$ term in the metric coincides
with that in the $\wt J=0$ case as $r\to 0$.
This gives:\footnote{In order to be able to carry out this rescaling
we need $\wt J^2 < 64 r_0^3$. {}From \refb{ep5} it follows that this is
the condition $(Q_1-Q_5)Q_5 n> J^2$ that guarantees that the
original black hole solution is supersymmetric. This is also the
condition needed for the absence of closed time-like
curves\cite{9602065}.}
\ben \label{ebm3}
ds^2 &=& \psi^{-1}\left[dudv + \left(K +{\wt J^2 \over 64 r_0^4} r \psi^2
-{\wt J^2\over 32 r_0^2 r}\right)
\left(1-{\wt J^2\over 64r_0^3}\right)^{-1}dv^2\right.
\nn && \left. + \left({\wt J\over 4r}
 - {\wt J r \over 4 r_0^2}\psi^2\right)\,
 \left(1-{\wt J^2\over 64r_0^3}\right)^{-1/2}
\, (dx^4+\cos\theta d\phi)
\, dv\right] \nn
&& + \psi\left[{dr^2\over r} + r
\left\{(dx^4+\cos\theta d\phi)^2 + d\theta^2 + \sin^2\theta d\phi^2
\right\}\right]\, .
\een
Let us denote the difference between this metric and the non-rotating
black hole metric \refb{eoror} by $\Delta (ds^2)$. We have
\ben \label{ediffe}
\Delta (ds^2) &=&
\psi^{-1}\left[\left(K +{\wt J^2 \over 64 r_0^4} r \psi^2
-{\wt J^2\over 32 r_0^2 r}\right) \left(1-
{\wt J^2\over 64 r_0^3}\right)^{-1} - K\right] dv^2
\nn &&
+ \psi^{-1} \left({\wt J\over 4r}
- {\wt J r \over 4 r_0^2}\psi^2\right)
\, \left(1-{\wt J^2\over 64r_0^3}\right)^{-1/2}
\, (dx^4+\cos\theta d\phi)
\, dv\, .
\een
Expressing this in the new coordinate system \refb{newcoord}
we get
\be \label{enewco}
\Delta (ds^2) = -{\wt J^2\over 8 r_0^{3/2}}
\left(1-{\wt J^2\over 64 r_0^3}\right)^{-1} W^2 V^{-1} dV^2
+ \hbox{n.s.}\, ,
\ee
where n.s. denotes terms which are non-singular as $V\to 0$.
{}From the form of the $\wt J=0$ metric given in \refb{eregm} we see that
the singular term in \refb{enewco} can
be removed by a shift in the
$U$ coordinate of the form
\be \label{eushift}
U \to U+{\wt J^2\over 32 r_0^{5/2}}
\left(1-{\wt J^2\over 64 r_0^3}\right)^{-1} \ln (-V)\, .
\ee
This makes the metric non-singular.

The three form field strength in this new coordinate system takes
the form
\begin{eqnarray}
  \label{eq:24mod}
F^{(3)} &=& {r_0\over \lambda}\, \bigg
[ \sin\theta\, 
dx^4\wedge d\theta\wedge d\phi + 4
WdW\wedge dV\wedge  dU   \nn
&& -  \frac{\tilde J}{r_0}\, \left(1-\frac{{\tilde J}^2}
{64 r_0^3}\right)^{-1/2}\,
WdW\wedge dV\wedge
  (dx^4 + \cos\theta
  d\phi)   \nonumber \\
&&  - \frac{\tilde J}{2r_0}\left(1-\frac{{\tilde J}^2}
{64 r_0^3}\right)^{-1/2}W^2 \, \sin\theta\,  
dV\wedge d\theta\wedge d\phi \bigg]\,  .
\end{eqnarray}

The analysis for $\wt J\ne 0$ black hole in Taub-NUT proceeds along
similar lines. The only difference is in the final step -- the shift in
$U$ given in \refb{eushift} is now replaced by
\ben \label{enewshift}  
U &\to& U+{\wt J^2\over 32 r_0^{5/2}}
\left(1-{\wt J^2\over 64 r_0^3}\right)^{-1} \ln (-V)
+ 4\, r_0^{3/2}\, R_4^{-2}  
\, \ln(-V)  \nn && 
- {3\wt J^2\over 16 r_0^{3/2} 
R_4^2} \left(1-{\wt J^2\over 64 r_0^3}\right)^{-1}
\ln (-V)\, .
\een

\sectiono{Regularity of the Deformed Solution} \label{sappc}

In this appendix we shall check whether the deformations we have
obtained by turning on various modes in \S\ref{s2} and \S\ref{s3}
produce regular field configuration at the future horizon. 
Since for deformation of the type we are considering  
the possible singularities are null singularities, they will not show
up in the invariant scalars constructed out of the field strengths
and the Riemann tensor. Instead
we need to work in a coordinate system in which the metric and the
other background fields are continuous at the horizon, and then check
whether the components of the Riemann tensor and other field
strengths are finite in this coordinate system\cite{9701077}.
We will
systematically carry out this analysis for all the modes, but not in
the same order in which they were analyzed in the text.  As in the
previous appendix, we will be working with the $\tilde J=0$ solution
for the sake of simplicity, but the generalization to $\wt J\ne 0$
case is straightforward.  We shall however continue to refer to the
$\wt J=0$ black holes as BMPV black holes.

We will start with the deformations described
in \S\ref{s3.2}. These are generated by the 2-form fields
in BMPV black hole in Taub-NUT space. The deformation in the 2-form
field given in \refb{enewf} near the origin behaves as
\be
\label{ehdefnew} \delta H^s \simeq{1\over R_4^2} h^s dv\wedge \left[
-{r } \sin\theta d\theta d\phi + dr\wedge (dx^4 +\cos\theta
d\phi)\right]\, . \ee
 In the new coordinate system it takes the form
\be \label{ehinnew} \delta\,
H^s \simeq{4 \, r_0^2\over R_4^2} h^s \,
dV\wedge \left[-W^2 \, \sin\theta\, d\theta \wedge d\phi + 2\, W\,
dW\wedge (dx^4 +\cos\theta \, d\phi)\right]\, . \ee
 This is clearly non-singular near the horizon $V=0$.
 Let us now examine the metric deformation
generated by these modes as given in eq.(\ref{enorm}):
\begin{equation}
  \label{eq:16}
  \psi^{-1} \, \wt S(v,\vec y)\, dv^2 = \psi^{-1} \,
  \frac{C(v)\, r}{2R_4^2(4r+R_4^2)}\, dv^2.
\end{equation}
This term in the new coordinates becomes
\begin{equation}
  \label{eq:14}
  \psi^{-1} \, \wt S(v,\vec y)\, dv^2
  = C(v) \, dV^2\, \frac{8 r_0^3 W^4 }{R_4^2
\left\{R_4^2 (1 + 4r_0^{1/2} V W^2 ) - 16 r_0^{3/2} V W^2 \right\}}
  \, .
\end{equation}
This is also regular at the horizon $V=0$.

Note however that since $v\propto \ln(-V)$ for small $|V|$, $v$ is a
rapidly varying function of $V$ near the horizon. Since $C(v)$ is an
oscillatory function of $v$ with finite period (set by the period of
the $x^5$ coordinate) $C(v)$ and hence $G_{VV}$ is a rapidly varying
oscillatory function of $V$ for small $V$. This can be remedied by a
shift in the $U$ coordinate of the form:
\begin{equation}
  \label{eq:19}
  U \to U - F(V,W), \qquad F(V,W)\equiv
  \frac{2r_0^2W^2}{R_4^4}\int_0^V C(v')dV'\, .  
\end{equation}
In this coordinate system $G_{VV}$ vanishes at $V=0$, but we get an
additional term in the $dWdV$ component proportional to $4\,
r_0W^2\p_W F(V,W)$.  Since $F(V,W)$ vanishes at $V=0$ for all $W$,
this additional term vanishes at the horizon.  Thus it does not alter
the behaviour of the metric at the horizon. We note however that $\p_V
F\propto W^2 C(v')$ is rapidly oscillating and as a result $\p_V^2 F$
diverges at the horizon. This could give a potential divergence in the
Riemann tensor which involves two derivatives of the metric. However
it can be seen using the argument below \refb{elaplace} that the
Riemann tensor never involves $\p_V^2 G_{WV}$. The latter term is of
weight 3 (we now replace $(u,v)$ by $(U,V)$ in counting weight)
whereas the Riemann tensor, written in a covariant form can have at
most two indices set equal to $V$ and hence can at most be of weight
2.  Thus $\p_V^2 G_{WV}$ cannot appear in the expression for the
Riemann tensor and the latter is finite at the horizon.

Next we look at the gravitino modes.  For definiteness we shall
consider the case of BMPV black holes in flat transverse space, but an
identical analysis can be carried out for Taub-NUT space.  The
gravitino modes in the original metric are non-vanishing only for the
$v$ components and these components take the form
\begin{equation}
  \label{eq:20}
  \Psi_v = \psi^{-3/2}(r)\eta(v,\theta,\phi), \qquad
  (\wt\Gamma^0 + \wt\Gamma^1)\eta(v,\theta,\phi) = 0\, ,
  \qquad \wh\Gamma^1 \eta(v,\theta,\phi)=-\eta(v,\theta,\phi)\, ,
\end{equation}
where $\eta(v,\theta,\phi)$ is an SO(5,1) spinor and also an $SO(5)$
spinor. The $(\theta,\phi)$ dependence of $\eta(v,\theta,\phi)$ was
computed in \S\ref{s2.2} and the $v$ dependence is arbitrary except
for the periodicity requirement imposed by the period of the
coordinate $x^5$.  In the new coordinate system the gravitino field
takes the form
\begin{equation}
  \label{eq:21}
  \Psi_V = (4\, r_0)^{5/4}\, W^3 (-2V)^{1/2}\,\eta(v,\theta,\phi)\, .
\end{equation}
This however is not the end of the story. The gravitino field
configuration \refb{eq:20} was computed using the set of vielbeins
\refb{eoneform} which become singular near the horizon.  This can be
seen by expressing them in the new coordinate system:
\ben \label{eorigv}
e^+ &\equiv& e^0 + e^1 = -r_0^{1/2}\, {dV\over V}\, , \nonumber \\
e^- &\equiv & e^1 - e^0 = -4 \, r_0^{1/2}\, V\, W^2 \, dU +
4 \, r_0^{1/2} \,
{dW\over W} +\left({r_0^{1/2}\over V} - 12\, r_0 W^2\right) \, dV \nonumber \\
e^3 &=& 2\, r_0^{1/2} \, (1+4 \, r_0^{1/2} \, V\, W^2)^{-3/2}
\left({dW\over W} + {dV\over 2V}
\right)\, .
\een
They are clearly singular at $V=0$.  Thus we must make a local Lorentz
transformation to make them non-singular.  {}From the metric
\refb{eregm} we see that a non-singular choice of vielbeins will
correspond to
\ben \label{ensv}
\tilde e^+ &=& 2\, r_0^{1/2} \, dV, \nonumber \\
\tilde e^- &=& 2 r_0^{1/2} \left[ W^2 dU + 8 r_0 W^4 Z^{-3}
(3 + 16 r_0^{1/2}\, V\, W^2
+ 24 r_0 V^2 W^4) dV \right. \nonumber \\ && \left.
- 4 r_0^{1/2} W Z^{-3} ( 3 + 12 r_0^{1/2} V W^2 +
16 r_0 V^2 W^4) dW \right], \nonumber \\
\tilde e^3 &=& 2 r_0^{1/2} W^{-1} Z^{-3/2} dW\, .
\een
The metric can be expressed as
\be \label{emetnew}
ds^2 = e^+ e^- + (e^3)^2 + (e^2)^2 + (e^4)^2 + (e^5)^2
=\tilde e^+ \tilde e^- + (\tilde e^3)^2 + (e^2)^2 + (e^4)^2 + (e^5)^2
\, .
\ee
We must now find the Lorentz transformation relating the two sets
of vielbeins. This is done in two steps. First we apply a boost
on $(e^+, e^-)$ that produces a new set of vielbeins:
\be \label{eboost}
\hat e^+ = -2\, V \, e^+, \quad \hat e^- = -{1\over 2\, V} e^-,
\qquad \hat e^3 = e^3\, .
\ee
The vielbeins $(\hat e^\pm, \hat e^3)$ can now be shown to be related
to $(\tilde e^\pm, \tilde e^3)$ by a Galilean transformation:
\be \label{egal}
\tilde e^+ = \hat e^+, \qquad \tilde e^- = \hat e^- - 2\, \beta\,
\hat e^3 - \beta^2 \, \hat e^+, \quad \tilde e^3 =\hat e^3 +
\beta\, \hat e^+\, ,
\ee
where
\be \label{edefbeta}
\beta = -{1\over 2\, Z^{3/2} \, V}\, .
\ee
These local
Lorentz transformation will also act on the gravitino fields.
First of all the boost transformation \refb{eboost} transforms
the gravitino to
\be \label{enewgrav}
\wh \Psi_V = (-2V)^{-1/2} \, \Psi_V =
(4\, r_0)^{5/4}\, W^3 \,\eta(v,\theta,\phi)\, .
\ee
The simplest way to see this
 is to note that the under this boost we must
have
\be \label{emust}
\overline{\Psi}_V \wt\Gamma^- {\Psi}_V\, e^+ = \overline{\wh\Psi}_V
\wt\Gamma^- \wh\Psi_V\, \hat e^+\, , \qquad \wt\Gamma^\pm
\equiv (\wt\Gamma^1\pm \wt\Gamma^0)\, ,
\ee
where we have used the $\wt\Gamma^+\Psi_V=0$ condition to infer that
the $\overline{\Psi}_V \wt\Gamma^m {\Psi}_V$ is non-vanishing
only for $m=-$.
Since $\hat e^+ = -2\, V \, e^+$ this implies that
$\wh \Psi_V = (-2V)^{-1/2} \, \Psi_V$.
On the other hand the Galilean transformation \refb{egal}
does not act on $\Psi_V$
since it is generated by $\wt\Gamma^{3+}$
and $\wt\Gamma^+$ annihilates
$\Psi_V$. Thus the gravitino in the frame given in \refb{ensv}
takes the form:
\be \label{efingrav}
\wt \Psi_V   =\wh\Psi_V =
(4\, r_0)^{5/4}\, W^3 \,\eta(v,\theta,\phi)\, .
\ee
Although this does not vanish at the horizon, we can make it vanish
using a local supersymmetry transformation by a parameter
proportional to $W^3 \,\int_0^V \, \eta(v',\theta,\phi)\, dV'$.

We note in passing that in this coordinate system the Killing spinor
behaves as $W\, V^{1/2}$ near $V=0$. After the local Lorentz transformation
described above it goes as $W$ and hence is well defined
at the horizon. We also note that the second solution given in
\refb{erew6}, for which $\Psi_v\sim r^{1/2}$ near $r=0$, diverges as
$V\to 0$ in the new coordinate system. Hence it is not an
allowed deformation.

Let us now
look at the modes generated by the
Garfinkle-Vachaspati transformation of the original
black hole metric.
We begin with the centre of mass motion modes of the four
dimensional black hole as described in \S\ref{s3.1}.
These modes are described by the metric
perturbation
\be \label{eyini}
\psi^{-1}\, g_i(v)y^idv^2=\psi^{-1} \, r\, n^i  g_i(v) \, dv^2, \qquad
  \vec n\equiv (\sin\theta\cos\phi,
\sin\theta\sin\phi, \cos\theta)\, .
\ee
In the new coordinate system this deformation is given by
\begin{equation}
  \label{eq:23}
 16 r_0^3 n^i g_i(v) (1 + 4\sqrt{r_0}\, V W^2)^{-1}\,
 W^4dV^2 \, .
\end{equation}
As in the case of the deformation
\refb{eq:14},
\refb{eq:23} takes finite value on the horizon but oscillates rapidly
as $V\to 0$, We can make $\delta G_{VV}$ vanish
by the coordinate transformation
\be \label{egtrs}
U\to U - H(V,W,\vec n), \qquad H(V,W,\vec n)=
4\, r_0^2 \, W^2\,
\int_0^V dV' \, n^i g_i(v') (1 + 4\sqrt{r_0}V' W^2)^{-1}\, .
\ee
This generates a term   $-4r_0 W^2(\p_W H  dV dW+\p_\theta H
dV d\theta +\p_\phi H dV d\phi)$ in the metric
but this vanishes at $V=0$ since $H(V,W,\vec n)$
vanishes at $V=0$ for
all $W$, $\theta$, $\phi$.\footnote{$H$ depends on $\theta$, $\phi$
through $\vec n$.}
$\p_V^2 H$ diverges at $V=0$, but as argued before
$\p_V^2 G_{WV}$, $\p_V^2 G_{\theta V}$ and
$\p_V^2 G_{\phi V}$ do not appear in the Riemann tensor for the type
of metric we are considering. Thus the metric and the Riemann tensor
are non-singular at $V=0$ for this deformation.

Finally we shall
carry out this analysis for the transverse oscillation of the
BMPV metric in the flat space, described in \S\ref{s2.1}.
The same analysis also holds for the modes describing the
transverse oscillation of the BMPV black hole relative to the
Taub-NUT space as described in \S\ref{s3.4} since they are
expected to have the same form near the horizon.
In this case the deformation is given by:
\be \label{egv1}
\delta(ds^2) = \psi^{-1} \, \vec f(v)\cdot \vec w\, dv^2 =
2\, r^{1/2}\, \psi^{-1} \, \vec f(v)\cdot \vec m\, dv^2,   
\qquad
\vec m=\vec w/|\vec w|\, .
\ee
In the new coordinate system this deformation takes the form
\be \label{egv2}
\delta(ds^2) = 16\, { r_0^{9/4} W^3 \over \sqrt{1 + 4\sqrt{r_0}
W^2 V}} {dV^2\over \sqrt{-V}} \, \vec m\cdot \vec f(v)\, .
\ee
The metric is singular at $V=0$. However this term can be removed
by the following shift of $U$:
\be \label{egv3}
U\to U - G(V,W,\vec m), \quad G(V,W,\vec m)
= 4\, r_0^{5/4} \, W \, \int_0^V dV' \, (1 + 4 \sqrt{r_0} W^2 V')^{-1/2}
\, (-V')^{-1/2} \, \vec m\cdot \vec f(v')\, .
\ee
This shift however generates a term in the metric of the form
\be \label{enewterm}
-4\, r_0\, W^2\, \p_W\, G(V,W,\vec m)\, dW \, dV
-4\, r_0\, W^2\, \p_i\, G(V,W,\vec m)\, d\theta^i \, dV\, ,
\ee
where $\theta^i$ denotes any of the angular coordinates.
These vanish at the horizon, but their first $V$ derivatives
diverge at the horizon as $\vec f(v) / (-V)^{1/2}$. This could give
rise to divergences in the Riemann tensor. Explicit computation
shows that some of the components of the Riemann tensor do indeed
diverge at $V=0$. For example we find
\be \label{eexample}
R_{VWVW} = -2 r_0\, \, W^{-1} \p_W \p_V (W^3   \p_W
G(W,V)) \,+\,\hbox{n.s.}
=  -24\, r_0^{9/4}\, W  \, (-V)^{-1/2}\, \vec m\cdot \vec f
\, +\, \hbox{n.s.}\, ,
\ee
where n.s. denotes non-singular terms.
This diverges as $V\to 0$. Thus we conclude that these modes should
not be counted among the hair degrees of freedom of the black hole.


\begin{thebibliography}{99}

  \bibitem{9307038}
  R.~M.~Wald,
  ``Black hole entropy in the Noether charge,''
  Phys.\ Rev.\ D {\bf 48}, 3427 (1993)
  {\tt arXiv:gr-qc/9307038}.

\bibitem{9312023}
  T.~Jacobson, G.~Kang and R.~C.~Myers,
  ``On Black Hole Entropy,''
  Phys.\ Rev.\ D {\bf 49}, 6587 (1994)
  {\tt arXiv:gr-qc/9312023}.

\bibitem{9403028}
  V.~Iyer and R.~M.~Wald,
  ``Some properties of Noether
  charge and a proposal for dynamical black hole
  entropy,''
  Phys.\ Rev.\ D {\bf 50}, 846 (1994)
  {\tt arXiv:gr-qc/9403028}.

\bibitem{9502009}
  T.~Jacobson, G.~Kang and R.~C.~Myers,
  ``Black hole entropy in higher curvature gravity,''
  {\tt arXiv:gr-qc/9502009}.

\bibitem{0809.3304}
  A.~Sen,
  ``Quantum Entropy Function from AdS(2)/CFT(1) Correspondence,''
  {\tt arXiv:0809.3304 [hep-th]}.

\bibitem{9602065}
  J.~C.~Breckenridge, R.~C.~Myers, A.~W.~Peet and C.~Vafa,
  ``D-branes and spinning black holes,''
  Phys.\ Lett.\ B {\bf 391}, 93 (1997)
  {\tt arXiv:hep-th/9602065}.

\bibitem{0209114}
  J.~P.~Gauntlett, J.~B.~Gutowski, C.~M.~Hull, S.~Pakis and H.~S.~Reall,
  ``All supersymmetric solutions of
  minimal supergravity in five dimensions,''
  Class.\ Quant.\ Grav.\  {\bf 20}, 4587 (2003)
  {\tt arXiv:hep-th/0209114}.

\bibitem{0503217}
  D.~Gaiotto, A.~Strominger and X.~Yin,
  ``New connections between 4D and 5D black holes,''
  JHEP {\bf 0602}, 024 (2006)
  {\tt arXiv:hep-th/0503217}.

\bibitem{9608096}
  R.~Dijkgraaf, G.~W.~Moore,
  E.~P.~Verlinde and H.~L.~Verlinde,
  ``Elliptic genera of symmetric products and second quantized strings,''
  Commun.\ Math.\ Phys.\  {\bf 185}, 197 (1997)
  {\tt arXiv:hep-th/9608096}.

\bibitem{9607026}
R.~Dijkgraaf, E.~P.~Verlinde and H.~L.~Verlinde,
``Counting dyons in N = 4 string theory,''
Nucl.\ Phys.\ B {\bf 484}, 543 (1997)
{\tt arXiv:hep-th/9607026}.

\bibitem{0412287}
G.~L.~Cardoso, B.~de Wit, J.~Kappeli and T.~Mohaupt,
``Asymptotic degeneracy of dyonic N = 4 string states
and black hole
entropy,''
JHEP {\bf 0412}, 075 (2004) {\tt arXiv:hep-th/0412287}.

\bibitem{0505094}
  D.~Shih, A.~Strominger and X.~Yin,
  ``Recounting dyons in N = 4 string theory,''
  JHEP {\bf 0610}, 087 (2006)
  {\tt arXiv:hep-th/0505094}.

\bibitem{0506249}
D.~Gaiotto,
``Re-recounting dyons in N = 4 string theory,''
 {\tt arXiv:hep-th/0506249}.

\bibitem{0605210}
  J.~R.~David and A.~Sen,
  ``CHL dyons and statistical entropy function from D1-D5 system,''
  {\tt arXiv:hep-th/0605210}.

\bibitem{0708.1270}
 A.~Sen,
 ``Black Hole Entropy Function, Attractors and Precision Counting of
 Microstates,''
 {\tt arXiv:0708.1270 [hep-th]}.

\bibitem{0901.0359}
  N.~Banerjee, I.~Mandal and A.~Sen,
  ``Black Hole Hair Removal,''
  {\tt arXiv:0901.0359 [hep-th]}.

\bibitem{9612248}
  N.~Kaloper, R.~C.~Myers and H.~Roussel,
  ``Wavy strings: Black or bright?,''
  Phys.\ Rev.\  D {\bf 55}, 7625 (1997)
  {\tt arXiv:hep-th/9612248}.

\bibitem{9701077}
  G.~T.~Horowitz and H.~s.~Yang,
  ``Black strings and classical hair,''
  Phys.\ Rev.\  D {\bf 55}, 7618 (1997)
  {\tt arXiv:hep-th/9701077}.

\bibitem{9605224}
  G.~T.~Horowitz and D.~Marolf,
  ``Counting states of black strings with traveling waves,''
  Phys.\ Rev.\  D {\bf 55}, 835 (1997)
  {\tt arXiv:hep-th/9605224}.

\bibitem{9606113}
  G.~T.~Horowitz and D.~Marolf,
  ``Counting states of black strings with traveling waves. II,''
  Phys.\ Rev.\  D {\bf 55}, 846 (1997)
  {\tt arXiv:hep-th/9606113}.

\bibitem{9505116}
  R.~Brooks, R.~Kallosh and T.~Ortin,
  ``Fermion zero modes and black hole hypermultiplet with rigid
  supersymmetry,''
  Phys.\ Rev.\  D {\bf 52}, 5797 (1995)
  {\tt arXiv:hep-th/9505116}.

\bibitem{romans}
  L.~J.~Romans,
  ``Selfduality For Interacting Fields:
  Covariant Field Equations For
  Six-Dimensional Chiral Supergravities,''
  Nucl.\ Phys.\  B {\bf 276}, 71 (1986).

\bibitem{9712176}
  F.~Riccioni,
  ``Tensor multiplets in six-dimensional (2,0) supergravity,''
  Phys.\ Lett.\  B {\bf 422}, 126 (1998)
  {\tt arXiv:hep-th/9712176}.

\bibitem{9804166}
   S.~Deger, A.~Kaya, E.~Sezgin and P.~Sundell,
   ``Spectrum of D = 6, N = 4b supergravity on AdS(3) x S(3),''
   Nucl.\ Phys.\  B {\bf 536}, 110 (1998)
   {\tt arXiv:hep-th/9804166}.

\bibitem{vachaspati}
  D.~Garfinkle and T.~Vachaspati,
  ``Cosmic string traveling waves,''
  Phys.\ Rev.\  D {\bf 42}, 1960 (1990).

\bibitem{9511064}
  F.~Larsen and F.~Wilczek,
  ``Internal Structure of Black Holes,''
  Phys.\ Lett.\  B {\bf 375}, 37 (1996)
  {\tt arXiv:hep-th/9511064}.

\bibitem{9604134}
  F.~Larsen and F.~Wilczek,
  ``Classical Hair in String Theory I: General Formulation,''
  Nucl.\ Phys.\  B {\bf 475}, 627 (1996)
  {\tt arXiv:hep-th/9604134}.

\bibitem{9609084}
  F.~Larsen and F.~Wilczek,
  ``Classical hair in string theory. II: Explicit calculations,''
  Nucl.\ Phys.\  B {\bf 488}, 261 (1997)
  {\tt arXiv:hep-th/9609084}.

\bibitem{9512031}
  M.~Cvetic and A.~A.~Tseytlin,
  ``Solitonic strings and BPS saturated dyonic black holes,''
  Phys.\ Rev.\  D {\bf 53}, 5619 (1996)
  [Erratum-ibid.\  D {\bf 55}, 3907 (1997)]
  {\tt arXiv:hep-th/9512031}.



\end{thebibliography}
\end{document}